\documentclass[article]{aa}  

\usepackage{graphicx}
\usepackage{epstopdf}
\usepackage{txfonts}
\usepackage{amsmath,amsfonts,amssymb}
\usepackage{color}

\begin{document} 

\title{The Gaia-ESO Survey: impact of extra mixing on C and N abundances of giant stars}

   \author{N. Lagarde
          \inst{1}, 
         C. Reyl\'e\inst{1}, A. C. Robin\inst{1},  G. Tautvai\v{s}ien\.{e}\inst{2}, A. Drazdauskas\inst{2}, \v{S}. Mikolaitis\inst{2}, R. Minkevi\v{c}i\={u}t\.{e}\inst{2}, E. Stonkut\.{e}\inst{2}, Y. Chorniy\inst{2}, V. Bagdonas\inst{2}, A. Miglio\inst{3}, G. Nasello\inst{1}, G. Gilmore\inst{4}, S. Randich\inst{5}, T. Bensby \inst{6}, A. Bragaglia\inst{7},  E. Flaccomio\inst{8}, P. Francois\inst{9}, A.~J. Korn\inst{10}, E. Pancino\inst{5,11}, R. Smiljanic\inst{12},  A. Bayo\inst{13}, G. Carraro\inst{14}, M.~T. Costado\inst{15},  F. Jim\'enez-Esteban\inst{16,17}, P. Jofr\'e\inst{18}, S.L. Martell\inst{19}, T. Masseron\inst{20,21}, L. Monaco\inst{22}, L. Morbidelli\inst{23}, L. Sbordone\inst{24}, S.~G. Sousa\inst{25}, S. Zaggia\inst{26}}

   \institute{Institut UTINAM, CNRS UMR6213, Univ. Bourgogne Franche-Comt\'e, OSU THETA Franche-Comt\'e-Bourgogne, Observatoire de Besan\c con, BP 1615, 25010 Besan\c con Cedex, France. \\
              \email{nadege.lagarde@utinam.cnrs.fr}
 \and  Institute of Theoretical Physics and Astronomy, Vilnius University,  Saul\.{e}tekio av. 3, 10257 Vilnius, Lithuania
  \and  School of Physics and Astronomy, University of Birmingham, Edgbaston, Birmingham B15 2TT, UK 
  \and Institute of Astronomy, University of Cambridge, Madingley Road, Cambridge CB3 0HA, United Kingdom
  \and INAF - Osservatorio Astrofisico di Arcetri, Largo E. Fermi 5, 50125, Florence, Italy
  \and Lund Observatory, Department of Astronomy and Theoretical Physics, Box 43, SE-221 00 Lund, Sweden
  \and INAF - Osservatorio Astronomico di Bologna, via Gobetti 93/3, 40129, Bologna, Italy
  \and INAF - Osservatorio Astronomico di Palermo, Piazza del Parlamento 1, 90134, Palermo, Italy
  \and GEPI, Observatoire de Paris, CNRS, Universit\'e Paris Diderot, 5 Place Jules Janssen, 92190 Meudon, France
  \and Department of Physics and Astronomy, Uppsala University, Box 516, SE-751 20 Uppsala, Sweden
  \and Space Science Data Center - Agenzia Spaziale Italiana, via del Politecnico, s.n.c., I-00133, Roma, Italy
  \and Nicolaus Copernicus Astronomical Center, Polish Academy of Sciences, ul. Bartycka 18, 00-716, Warsaw, Poland
  \and Instituto de F\'isica y Astronomi\'ia, Universidad de Valparai\'iso, Chile
  \and Dipartimento di Fisica e Astronomia, Universit\`a di Padova, Vicolo dell'Osservatorio 3, 35122 Padova, Italy
  \and Departamento de Did\'actica, Universidad de C\'adiz, 11519 Puerto Real, C\'adiz, Spain
  \and Departmento de Astrof\'{\i}sica, Centro de Astrobiolog\'{\i}a (INTA-CSIC), ESAC Campus, Camino Bajo del Castillo s/n, E-28692 Villanueva de la Ca\~nada, Madrid, Spain
  \and Suffolk University, Madrid Campus, C/ de la Vi\~na 3, 28003, Madrid, Spain
  \and N\'ucleo de Astronom\'{i}a, Facultad de Ingenier\'{i}a, Universidad Diego Portales, Av. Ej\'ercito 441, Santiago, Chile
  \and School of Physics, University of New South Wales 2052 Sydney, Australia
  \and Instituto de Astrof\'{\i}sica de Canarias, E-38205 La Laguna, Tenerife, Spain   
  \and Universidad de La Laguna, Dept. Astrof\'{\i}sica, E-38206 La Laguna, Tenerife, Spain
  \and Departamento de Ciencias Fisicas, Universidad Andres Bello, Fernandez Concha 700, Las Condes, Santiago, Chile
  \and INAF - Osservatorio Astrofisico di Arcetri, Largo E. Fermi 5, 50125, Florence, Italy
  \and European Southern Observatory, Alonso de Cordova 3107 Vitacura, Santiago de Chile, Chile
  \and Instituto de Astrof\'isica e Ci\^encias do Espa\c{c}o, Universidade do Porto, CAUP, Rua das Estrelas, 4150-762 Porto, Portugal
  \and INAF - Padova Observatory, Vicolo dell'Osservatorio 5, 35122 Padova, Italy
                }

\date{Received ?/ Accepted ?}
\authorrunning{Lagarde et al.} \titlerunning{The Gaia-ESO survey: impact of extra mixing on C and N abundances of giant stars}
 
  \abstract 
   {The Gaia-ESO Public Spectroscopic Survey using FLAMES at the VLT has obtained high-resolution UVES spectra for a large number of giant stars, allowing a determination of the abundances of the key chemical elements carbon and nitrogen at their surface. The surface abundances of these chemical species are known to change in stars during  their evolution on the red giant branch (RGB) 
  after the first dredge-up episode, as a result of the extra mixing phenomena.}
  {We investigate the effects of thermohaline mixing on C and N abundances using the first comparison between the Gaia-ESO survey [C/N] determinations with simulations of the observed fields using a model of stellar population synthesis.}
     { We explore the effects of thermohaline mixing on the chemical properties of giants through stellar evolutionary models computed with the stellar evolution code STAREVOL. We include these stellar evolution models in the Besan\c con Galaxy model to simulate the [C/N] distributions determined from the UVES spectra of the Gaia-ESO survey and to compare them with the observations.}
   {Theoretical predictions including the effect of thermohaline mixing are in good agreement with the observations. However, the field stars in the Gaia-ESO survey with C and N abundance measurements have a metallicity close to solar, where the efficiency of thermohaline mixing is not very large. The C and N abundances derived by the Gaia-ESO survey in open and globular clusters clearly show the impact of thermohaline mixing at low metallicity, which explains the [C/N] value observed in lower mass and older giant stars. Using independent observations of carbon isotopic ratio in clump field stars and open clusters, we also confirm that thermohaline mixing should be taken into account to explain the  behaviour of $^{12}$C/$^{13}$C as a function of stellar age.} 
  {Overall, the current model including thermohaline mixing is able to reproduce very well the C and N abundances over the whole metallicity range investigated by the Gaia-ESO survey data.}
                  \maketitle
%

\section{Introduction}

Over the last decade, the understanding of our Galaxy has dramatically increased thanks to the development of large spectroscopic surveys that provide fundamental properties of a large number of stars in different regions of our Galaxy (e.g. RAVE \citep{RAVEDR1}, SEGUE  \citep{SEGUE}, Gaia-ESO \citep{GilmoreGES}, APOGEE \citep{APOGEE17}, LAMOST \citep{LAMOST}, GALAH  \citep{GALAH}. Some of them provide abundances deduced from  high-resolution spectra, allowing the determination of light chemical elements such as carbon, nitrogen, or lithium. \\

The carbon and nitrogen abundances, and also the carbon isotopic ratio, are key chemical tracers used  to constrain the stellar evolution of giant stars. Indeed, low-mass stars experience the well-known first dredge-up at the bottom of the red giant branch (RGB), implying changes of the surface abundances of C and N \citep{Iben67}. After this episode, numerous spectroscopic observations show that an extra mixing occurs after the bump luminosity on the red giant branch changing the abundances of elements lighter than oxygen at the surface of bright red giant stars \citep[e.g.][]{Gilroy89, GiBr91, Gratton00, Luck94, Tautvaisiene00, Tautvaisiene01, Tautvaisiene05, Smiljanic09, Mikolaitis10,Mikolaitis12}.\\

Different transport processes have been discussed in the literature to explain the abundance anomalies in giants. Parametric computations have been proposed to better understand the behaviour of chemical abundances at the stellar surface in low- and intermediate-mass stars. After showing that the hot bottom burning (HBB) process, which was previously proposed by \citet{CameronFowler1971} to allow Li production in AGB stars, can explain the oxygen isotopic ratios in AGB stars with initial stellar mass between 4.5 and 7.0 M, \citet{Boothroyd95} introduced the notion of cool bottom processing (i.e. ad hoc transport material
from the cool bottom of stellar convective envelope to deeper and hotter radiative regions where nuclear reactions occur) to explain surface abundances of giant stars with masses lower than 2.0 M$_\odot$ \citep[see also][]{Wasserburgetal1995, BoSa99, SaBo99}. \citet{DeWe96} suggested, as did  \citet{Wasserburgetal1995}, the presence of non-standard mixing of unknown physical origin, between the hydrogen burning and the base of convective envelope after the bump to explain abundances anomalies in red giant stars. To understand variations in surface abundances of giant stars, they proposed a deep diffusive mixing \citep[e.g.][]{DeWe96, Weiss96, Denissenkov98}. Nevertheless, these two propositions of extra mixing are then not related to any physical mechanism to explain changes in surface abundances, and depend on free parameters. On the other hand, rotation has been investigated as a possible source of mixing in RGB stars by several authors \citep{SwMe79,Charbonnel95,DeTo00,Palacios06,Chaname05}, showing that the total
transport coefficient of rotation at this phase is too low to imply abundance variation on the first ascent giant branch as requested by observations of RGB stars brighter than the RGB bump.

Thermohaline instability driven by $^3$He-burning through the pp-chain has been proposed to govern the photospheric compositions of bright low-mass red giant stars \citep[e.g.][]{ChaZah07a, ChaLag10}. This double diffusive instability, induced by a mean molecular weight inversion due to the $^3$He($^3$He,2p)$^4$He reaction in the thin radiative layer between the convective envelope and the hydrogen-burning shell \citep{Eggleton08, Lattanzio15}, is a  physical mechanism that best reproduces the observational abundances (e.g. of C and N) in giant stars \citep[e.g.][]{ChaLag10, Angelou11, Angelou12, Henkel17}. A few recent papers have suggested that magnetic fields might play a role in stellar mixing, alone \citep{Busso07} or in combination with thermohaline mixing \citep{DenissenkovMerryfield10,Palmerini11}. Since thermohaline also provides  a physical solution of the $^3$He problem, well-known in  Galactic chemical evolution models \citep{Lagarde11,Lagarde12b}, we  focus on this mechanism in this paper. \\

To exploit all the potential of the spectroscopic data, \citet{Lagarde17} (hereafter L17) improved the Besan\c con Galaxy model (hereafter BGM) including the stellar evolution models that provide surface chemical and seismic properties of stars during their life. In addition to global properties, the BGM is now able to explore the effects of extra mixing on the surface abundances of different chemical species. The BGM allows us to compute the stellar component of our Galaxy taking into account errors and the selection function of the observations, drawing a consistent picture of the Galaxy with the formation and evolution scenarios of the Milky Way, stellar formation and evolution theory, models of stellar atmospheres, and dynamical constraints \citep{Robin03, Czekaj14}. This population synthesis model is a powerful tool that can improve current stellar evolution models and the physics of different transport processes occurring in stellar interiors using a comparison with observations of field stars at different stellar masses, metallicities, ages, evolutionary stages, or at different locations in the Galaxy.\\

In this paper, we  study the effects of thermohaline mixing with metallicity and mass of giant stars, using the C and N  abundances derived for the giants in the Gaia-ESO survey. To this end, we perform simulations using the BGM with and without the effects of  thermohaline instability. The simulations and data used for this study are described in Sect. \ref{data}, while the theoretical effects of thermohaline instability with stellar mass and metallicity are discussed in Sect.\ref{cn}. We start with field stars observed with UVES in Sect. \ref{field}, and then enlarge our study to the open and globular clusters  observed by the Gaia-ESO survey and compiled from the literature (see Sect.\ref{OCGC}). We draw our conclusions in Sect. \ref{conclu}.

  \begin{figure*}[h]
   \centering
  \includegraphics[width=0.9\hsize,clip=true,trim= 1cm 1cm 1cm 1cm]   {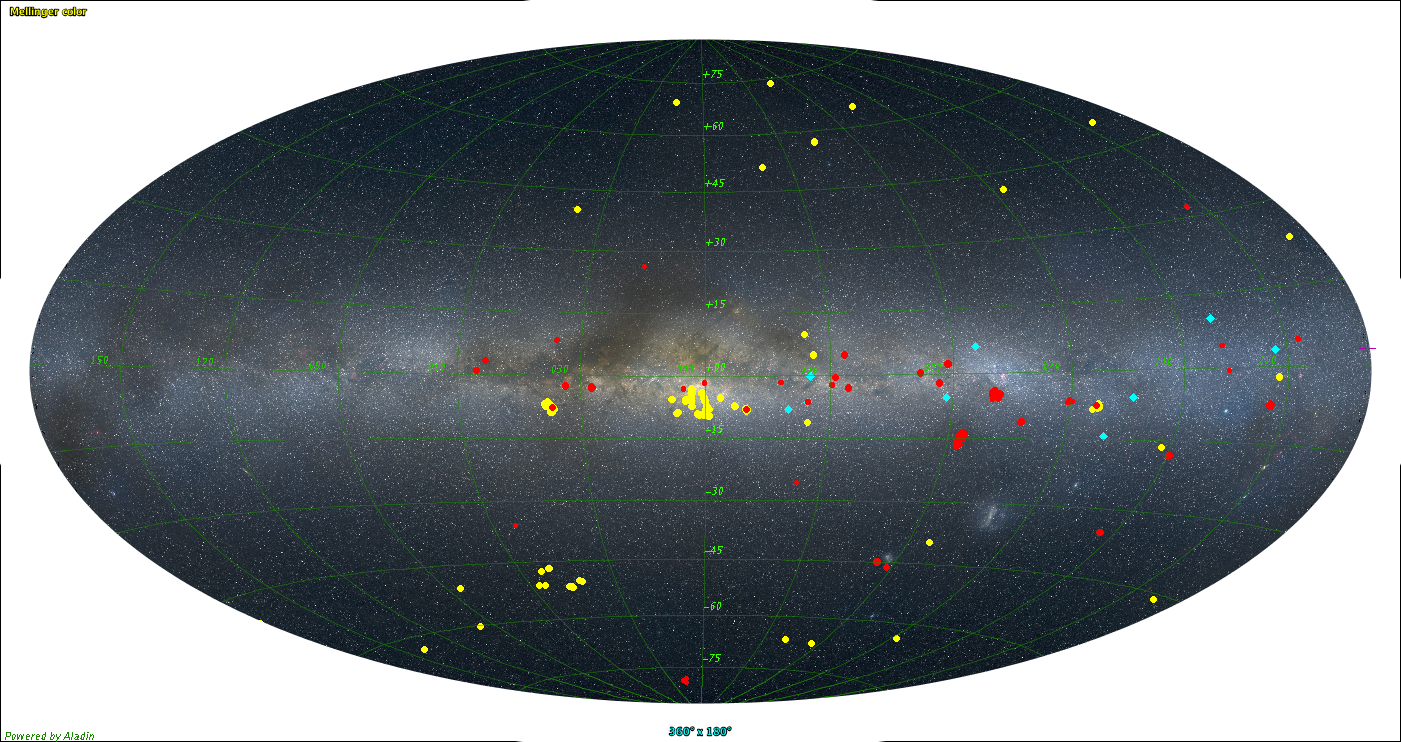}
   \caption{ All-sky view centred on the Milky Way Plane in an Aitoff projection from \citet{Mellinger09} with a Galactic coordinate grid. Giants observed by the Gaia-ESO survey for which C and N abundances are determined are colour-coded: clusters members (red) and field stars (yellow). We also represent nine open clusters (Collinder 261, Melotte 66, NGC6253, NGC3960, NGC2324, NGC2477, NGC2506, IC4651, NGC6134) for which $^{12}$C/$^{13}$C is derived (cyan diamond, see Sect. \ref{c1213_para}).}
   \label{planis}
 \end{figure*}


\section{Data set and simulations}
\label{data}

\begin{figure}
      \centering
   \includegraphics[width=8cm,clip=true,trim= 0cm 0cm 0cm 0cm]
   {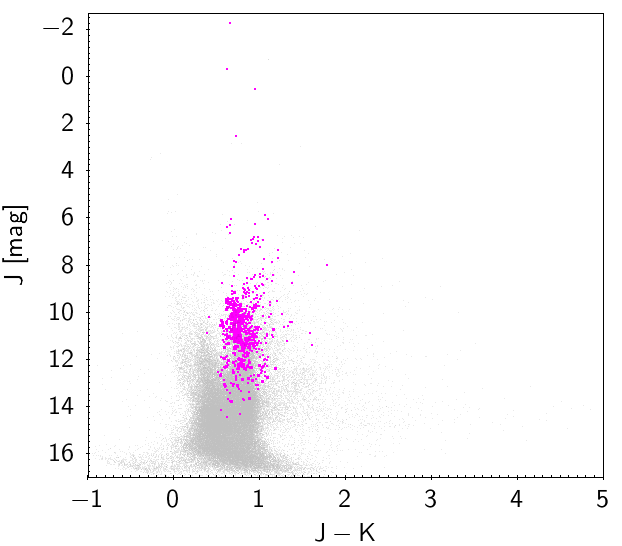}
         \caption{Colour-magnitude diagram for stars observed by the GES survey (grey dots), and for which C and N abundances have been determined (magenta dots). The colour and magnitude values are from the 2MASS Catalog.}
                  \label{colordiag}
   \end{figure} 

The Gaia-ESO survey \citep[][hereafter GES]{GilmoreGES,RandichGES} uses the  Fibre Large Array Multi Element Spectrograph (FLAMES) multifibre facility \citep{Pasquini02} of the Very Large Telescope (VLT) to obtain a better understanding of the kinematic and chemical evolution of our Galaxy. Giraffe, the medium-resolution spectrograph (R$\sim$20 000), and UVES \citep{Dekker00}, the high-resolution spectrograph (R$\sim$47 000), are used to observe up to 10$^5$ stars in the Milky Way. 

For our study, we used the observations of giant stars made with UVES, and the carbon and nitrogen abundances derived from
their spectra. These giant stars lie in different Galactic regions (see Fig. 1). All data used in this paper are included in the second, fourth, and fifth internal GES data releases (iDR2, iDR4, and iDR5) to have a sample that is as large as possible. The main atmospheric parameters of the stars were determined as described by \citet{Smiljanic14}, the carbon and nitrogen abundances were determined as described by \citet{Tautvaisiene15}. We separated the stars into two groups:
\begin{itemize}
\item 324 giant field stars with 173 stars located in the Galactic bulge;\item Giants belonging to open and globular clusters (see Table 1).
\end{itemize}

The simulations were made using the revised version of BGM \citep[Paper I of this series, ][]{Lagarde17}, where a new grid of stellar evolution models computed with the code STAREVOL (e.g. \cite{Lagarde12b, Amard16}) has been implemented. This new grid provides the global properties (e.g. surface gravity, effective temperature) and chemical abundances (for 54 stable and unstable species).
These models also take  into account the effects of thermohaline instability during the red giant branch \citep[e.g.][]{ChaLag10}. As discussed in Paper I, thermohaline instability changes the photospheric composition of low-mass brighter giant stars, with a decrease in carbon and an increase in nitrogen. In addition to the simulations discussed in \citet{Lagarde17}, improvements have been extended to all populations other than the thin disc, implying the computation of stellar evolution models with different $\alpha$-enhancements ([$\alpha$/Fe]=0.15 and 0.30), following the observational [$\alpha$/Fe] versus [Fe/H] trend observed by  Data Release 12 of APOGEE \citep{APOGEE15}. Namely, for $[Fe/H]<0.1$,
\begin{equation}
      [\alpha/Fe] =
     \left\{
     \begin{array}{rl}
     0.014+0.0140675\times[Fe/H]+0.101262\times[Fe/H]^{2} \\
      \text{for\ the\ thin\ disc\ stars}, \\
     0.320-exp(1.19375\times[Fe/H]-1.6) \\
    \text{for the thick disc stars}, \\
    0.3 \\
    \text{ for halo stars}.
     \end{array}
     \right.
     \end{equation}

For $[Fe/H]>$0.1, $[\alpha$/Fe] is assumed solar. To these relations an intrinsic Gaussian dispersion of 0.02 dex is added.

As discussed in \citet{Czekaj14}, the BGM also simulates the Poisson noise in the Monte Carlo generation of the simulated stars. We performed simulations in every GES field referred to in the iDR5 using this new version of the BGM. As shown in Figure~\ref{colordiag}, all stars in the sample have $0.5<J-K<1.0$ and $5.0<J<14$, so we restricted our simulations to these colour and magnitude ranges. 
We did two sets of simulations, with and without the effects of thermohaline instability. The selection bias introduced by the additional requirement of measurable carbon and nitrogen abundances cannot be taken into account in our sample, which is why simulations produce more stars than are present in the sample. This difference does not affect the conclusions of this paper.

\section{Thermohaline mixing effects on [C/N] }
\label{cn}
\subsection{Physics}

    \begin{figure}
      \centering
      \textbf{Low-RGB}\par
     \includegraphics[width=8cm,clip=true,trim=0cm 0cm 0cm 0cm]{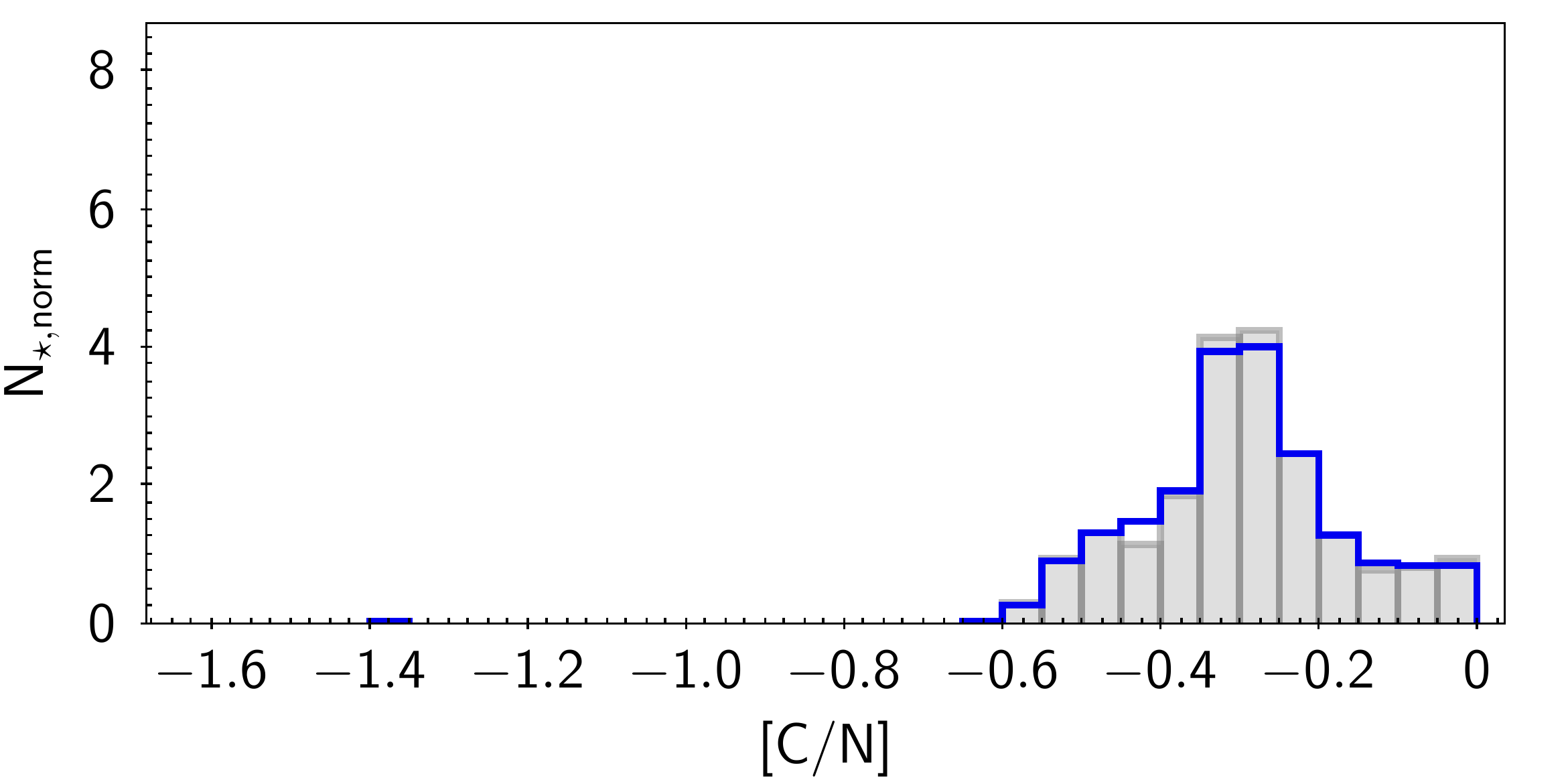}
      \par\vspace{0.3cm}
      \textbf{Upper-RGB}\par
 \includegraphics[width=8cm,clip=true,trim= 0cm 0cm 0cm 0cm]
   {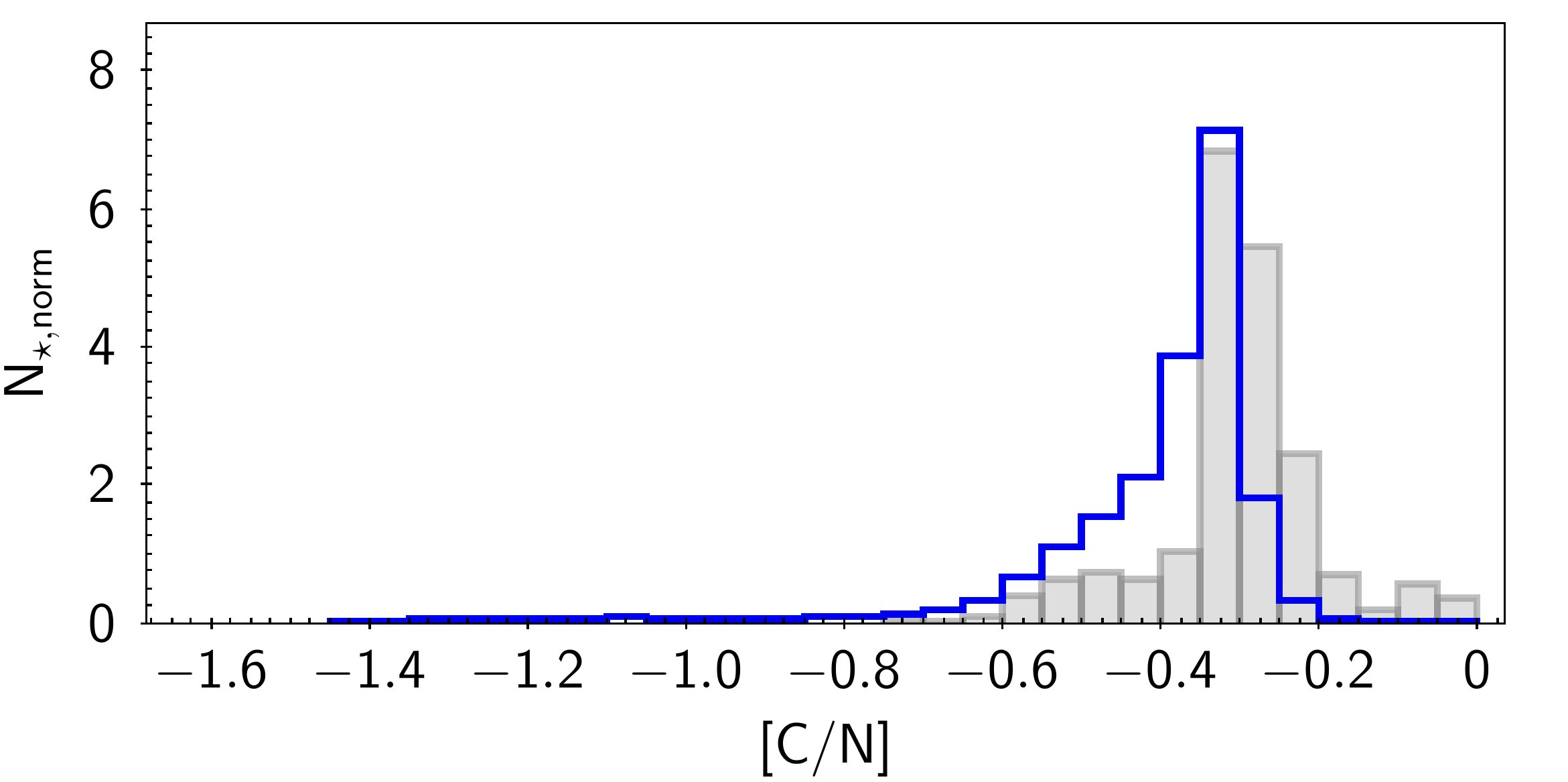}
      \par\vspace{0.3cm}
      \textbf{Clump/early-AGB}\par
   \includegraphics[width=8cm,clip=true,trim= 0cm 0cm 0cm 0cm]
   {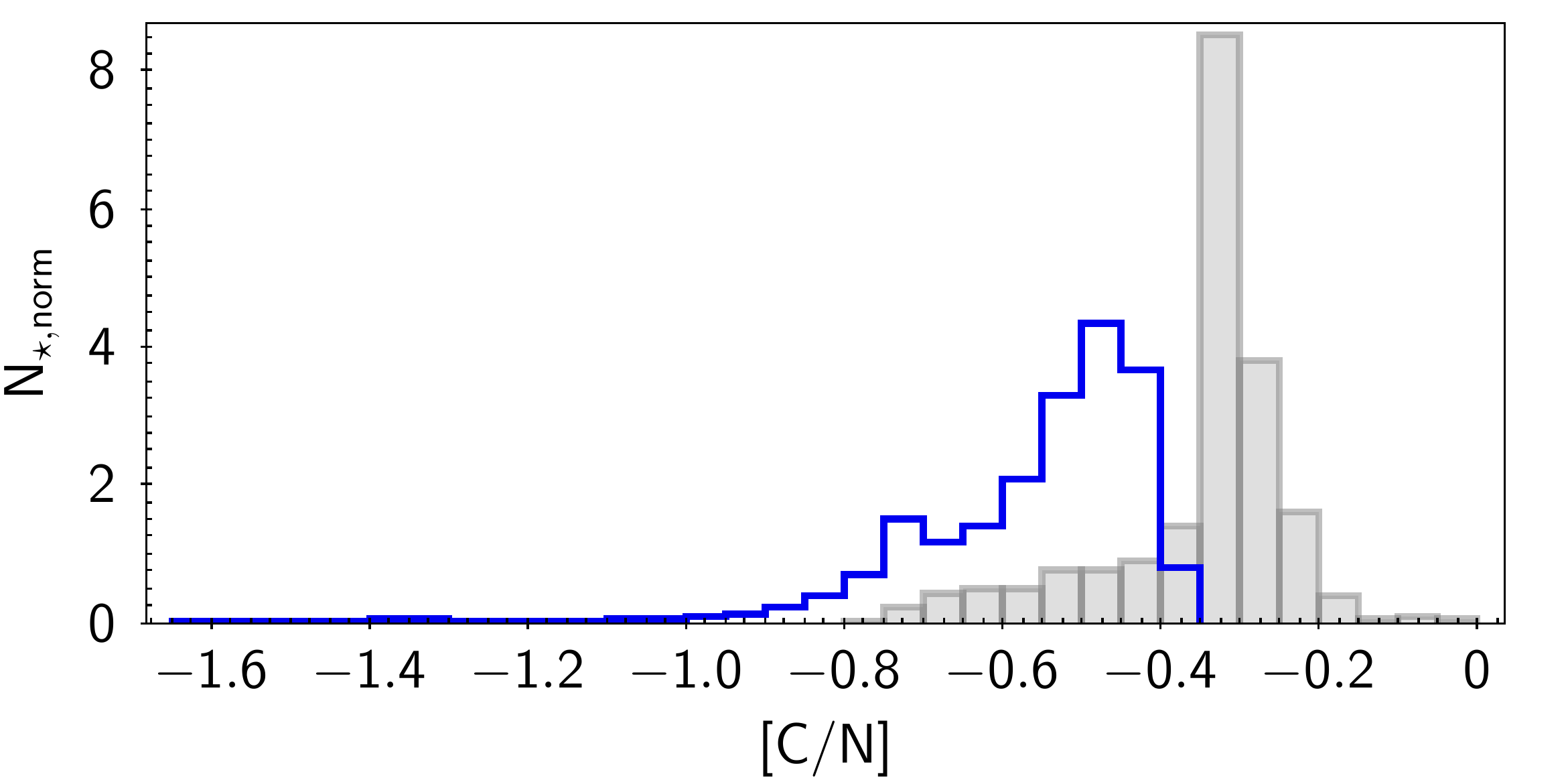}
         \caption{[C/N] distributions normalised to the histogram area of the synthetic population simulated with the BGM with the effects of thermohaline instability (blue solid histograms) and without (grey shaded histograms)  during the red giant branch. Giant stars are divided into three groups: low-RGB stars (before the RGB-bump luminosity, top panel), upper-RGB stars (after the RGB-bump luminosity, middle panel),  and clump/early-AGB stars (bottom panel). }
         \label{CNdistri}
   \end{figure}

Thermohaline mixing, as discussed in Paper I, is a double diffusive process conducted in RGB stars by an inversion of mean molecular weight induced by $^3$He($^3$He,2p)$^4$He reaction \citep{ChaZah07a,Siess09,ChaLag10, Henkel17} and a temperature gradient. For this study we use the prescriptions advocated by \citet{ChaZah07a} given by \citet{Ulrich72} with an aspect ratio of instability fingers $\alpha\sim$ 6 as referred by \citet{Krish03}. The thermohaline diffusion coefficient used in stellar evolution models includes the correction for non-perfect gas and is given by 

\begin{equation}
 D_ t =  {C_ t} \,  {K}  \left(\frac{\varphi} {\delta}\right)\frac{- \nabla_\mu }{(\nabla_{\rm ad} - \nabla)} \quad \hbox{for} \;  \nabla_\mu < 0, 
\label{dt}
\end{equation}
where $K$ is the thermal diffusivity; $\varphi = (\partial \ln\rho / \partial \ln\mu)_{P,T}$; $\delta=-(\partial \ln\rho/ \partial \ln \nu)_{P,\mu}$; $\nabla=(\partial \ln {\rm T}  / \partial \ln {\rm P})$; $\nabla_\mu$ and $\nabla_{\rm ad}$ are respectively the molecular weight gradient and  the adiabatic gradient; and with the non-dimensional coefficient
\begin{equation}
{C_ t} = \frac{8}{3} \pi^2 \alpha^2. 
\end{equation}

The value of $\alpha$ is still discussed in the literature by hydrodynamic simulations in 2D or 3D \citep{Denissenkov09, Denissenkov10, DenissenkovMerryfield10, RosenblumGaraudetal11, Traxleretal11, Brown13, GaBr15}. Although these simulations are still far from stellar conditions and do not take into account the coupling of this instability with other hydrodynamic processes occurring in red giant stars (e.g. rotation, magnetic field), they predict that thermohaline instability is not efficient enough to explain surface chemical abundances \citep{Wachlin14} with the thermohaline
fingers becoming more like blobs. Future hydrodynamical simulations closer to the conditions met in stellar interiors \citep{Prat15} or including the effects of other hydrodynamical processes as recently by \citet{SenGar18} would shed light on this discrepancy. Hence, we choose to use the prescriptions described above to compare 
theoretical predictions with current observations at different masses and metallicities. \\

\subsection{Impact along the evolution}

This double diffusive instability develops starting from the luminosity of the bump during the RGB \citep[e.g.][and Paper I]{ChaLag10}. Indeed, the steady increase in the stellar luminosity along the RGB momentarily stops when the hydrogen burning shell (HBS) crosses the molecular weight barrier left behind by the first dredge-up. At that moment the mean molecular weight of the HBS becomes smaller, which implies a decrease in the total stellar luminosity. This is called a bump in the luminosity function. When the region of nuclear energy production has passed this discontinuity, the mean molecular weight slightly increases and the stellar luminosity increases again. This variation in the luminosity causes an accumulation of stars in the colour-magnitude diagram leading to a bump in the luminosity distribution \citep[e.g.][]{Iben67, Iben68,Fusipecci90,Charbonnel94,Christensen15}.
The impact of thermohaline instability on the theoretical [C/N]\footnote{[X/Y]=A(X)$-$A(X)$_\odot-$A(Y)+A(Y)$_\odot$, with A(X)=log(N(X)/N(H))+12}  distributions at different evolutionary states are shown in Fig. \ref{CNdistri}.  With stellar models, giant stars are divided into three groups: (1) low-RGB stars: stars ascending the red giant branch before the RGB bump\footnote{Since the gravity (and thus luminosity) of the RGB-bump changes with the metallicity of stars, we establish a simple empirical relation based on stellar evolution models (without the effects of rotation on the evolutionary path) allowing the distinction of low-RGB and upper-RGB stars: log g$_{RGBbump}$=0.32$\cdot$[Fe/H]+2.44}. These stars have not yet undergone thermohaline mixing; (2) upper-RGB:  RGB stars brighter than the RGB-bump (with logg$\lesssim$2.2) ; (3) clump stars selected according to their asymptotic period spacing of g-modes $\Delta\Pi_{\ell=1}$. As discussed by \citet{ChaLag10} and Paper I, thermohaline instability occurring at the RGB-bump luminosity changes the surface abundances of C and N for giant stars brighter than the RGB-bump and along the red giant branch, resulting in a decrease in [C/N]  (see middle panel of Fig. \ref{CNdistri}). While thermohaline mixing is no longer happening in red clump stars, they have the lowest [C/N] (as shown in Fig. \ref{CNdistri}) because they have undergone a full RGB phase of extra mixing.\\

\subsection{Impact as a function of stellar mass}

As discussed and explained by \citet{ChaLag10} (see \S3.1.2), the global efficiency of thermohaline mixing increases when  considering less massive stars at a given metallicity or more metal-poor stars at a given stellar mass. This results from the combination of several factors like the thermohaline diffusion timescale compared to the secular timescale, the compactness of the hydrogen burning shell and of the thermohaline unstable region, and the amount of $^3$He available to power the thermohaline instability.\\

      \begin{figure}
   \centering
      \textbf{M $= 1.0$ M$_\odot$}\par
   \includegraphics[width=8cm,clip=true,trim= 0cm 0cm 0cm 0cm]
   {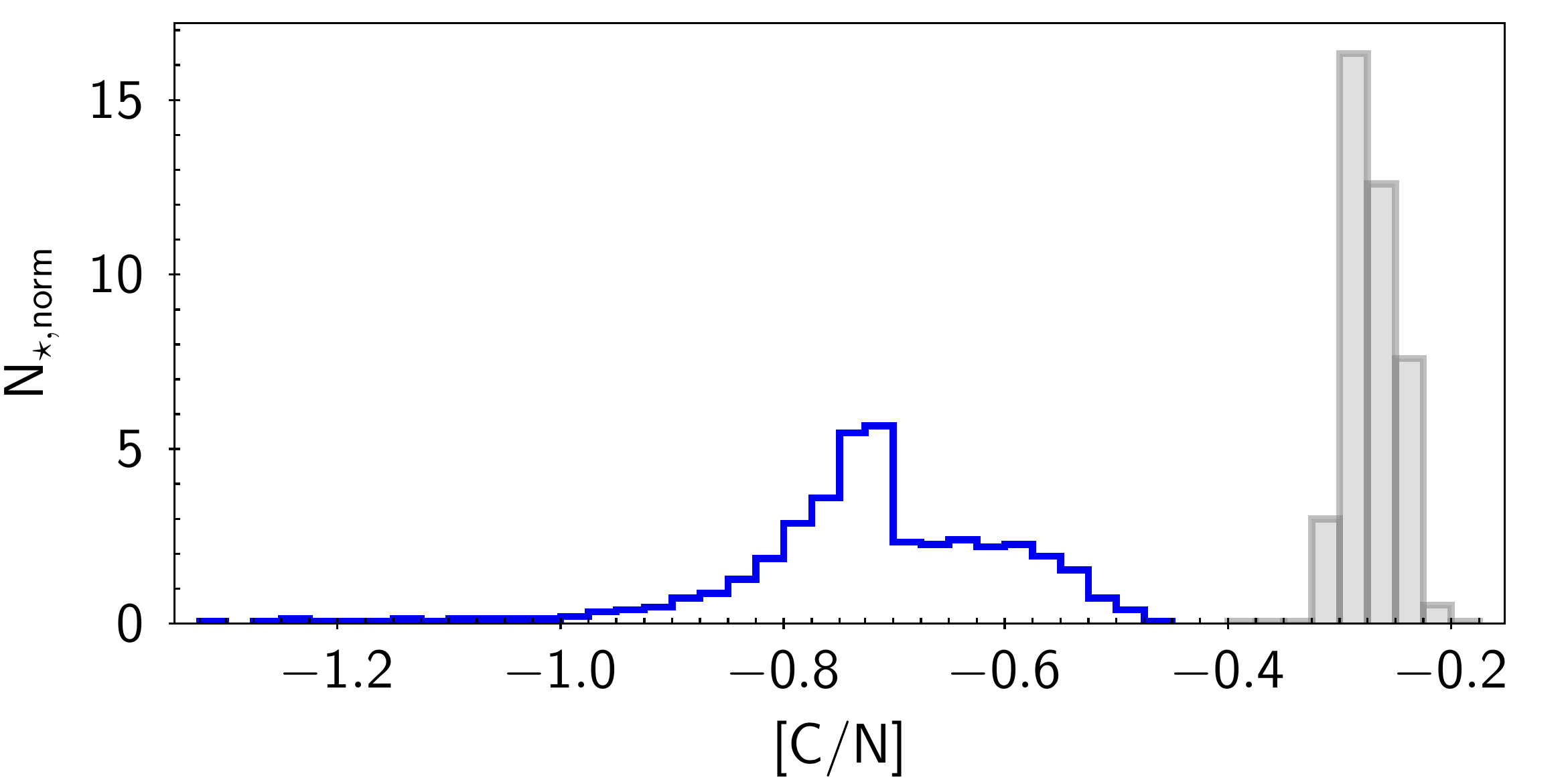}
      \par\vspace{0.3cm}
       \textbf{M $= 2.0$ M$_\odot$}\par
   \includegraphics[width=8cm,clip=true,trim= 0cm 0cm 0cm 0cm]
   {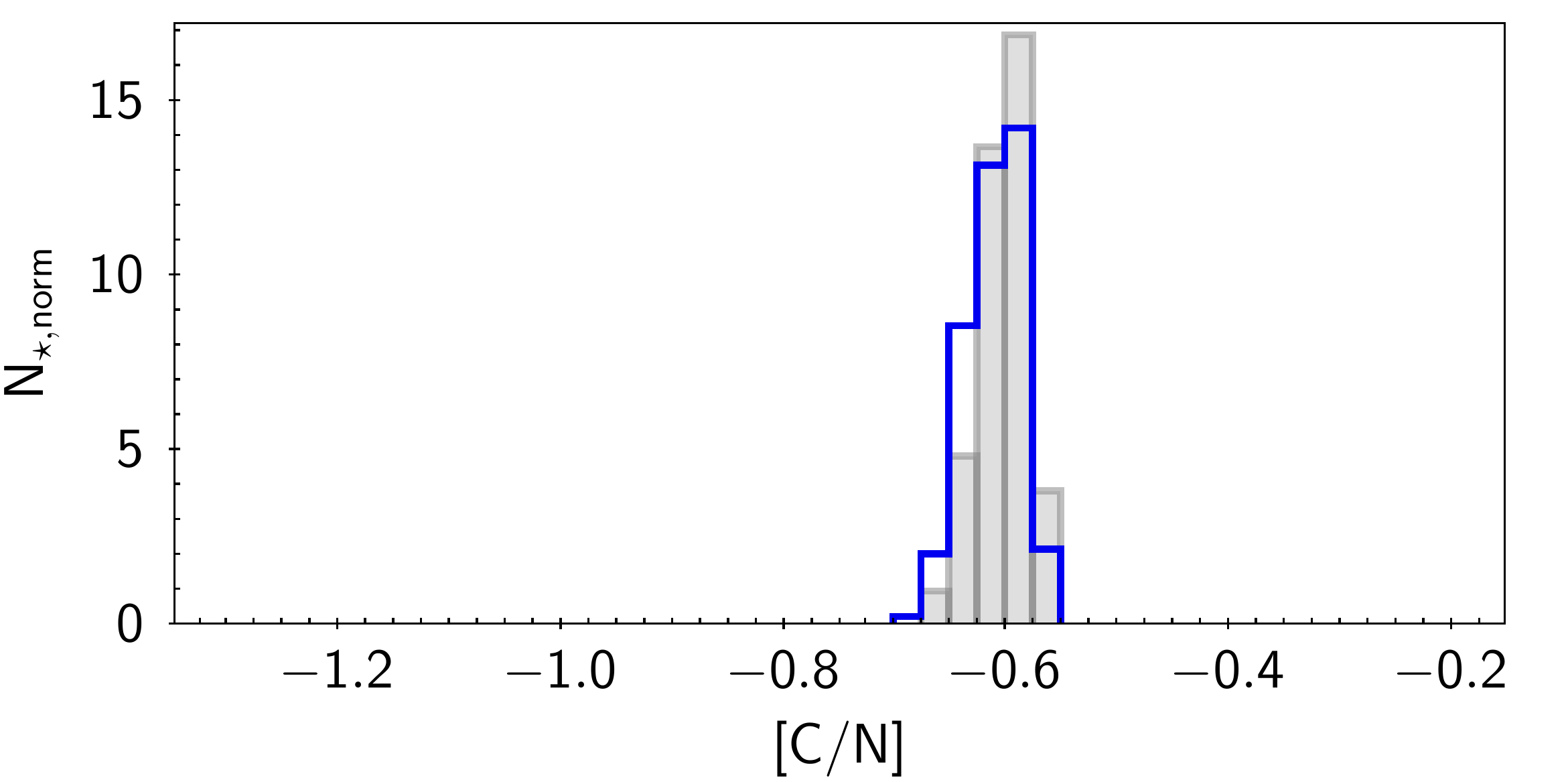}
      \par\vspace{0.5cm}
    \includegraphics[width=8cm,clip=true,trim= 0cm 0cm 0cm 0cm]
   {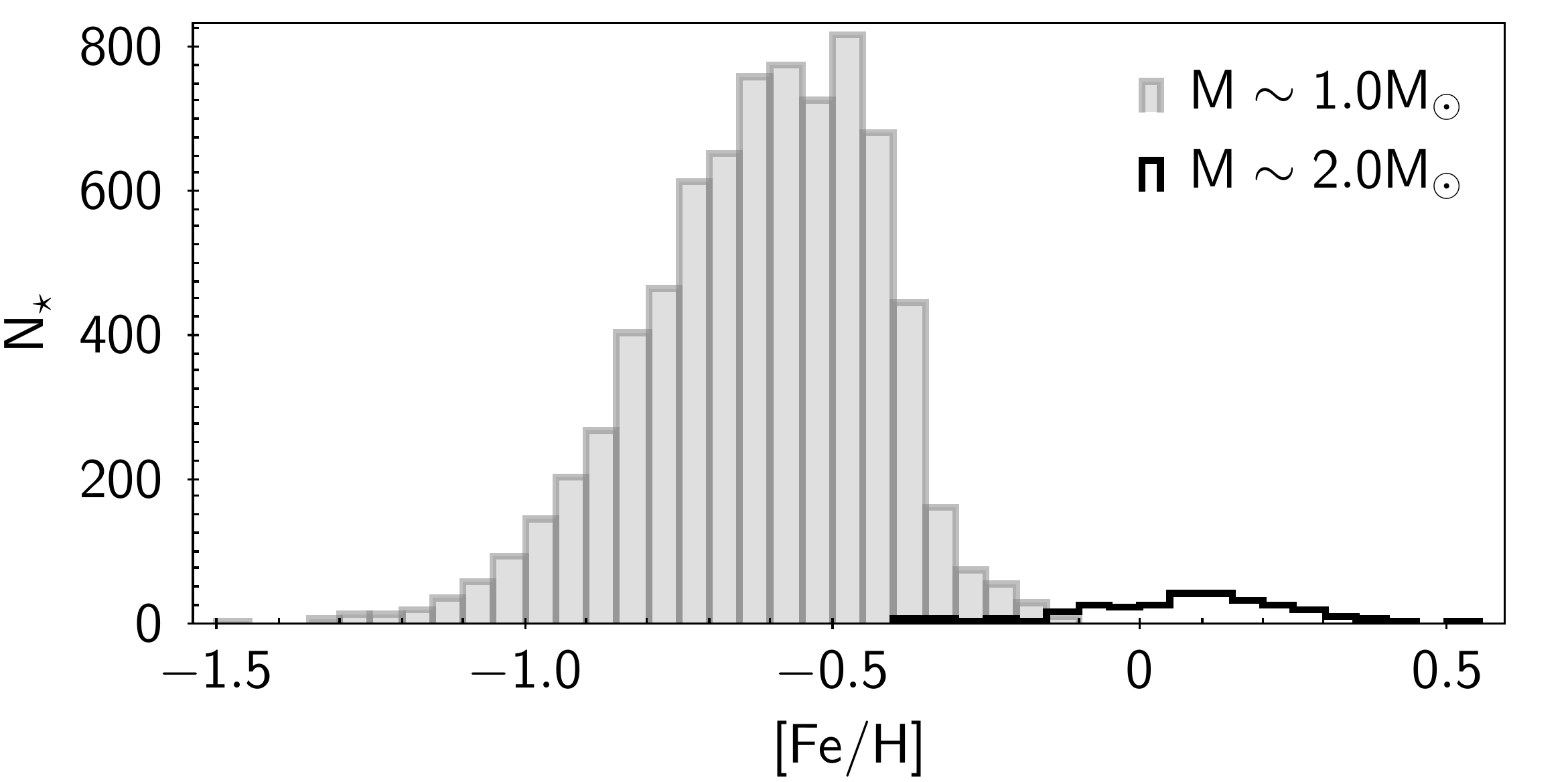}
      \caption{[C/N] distributions for two stellar masses ranges, 0.95$\leq$M/M$_{\odot}$$\leq$1.05 and 1.95$\leq$M/M$_{\odot}$$\leq$2.05 (top and bottom panel, respectively), for clump stars simulated with the BGM with the effects of thermohaline instability (blue solid histograms) and without  (grey shaded histograms). The bottom panel presents the mass distributions for clump stars simulated with the BGM at both mass ranges.}
         \label{CNdistriMass}
   \end{figure}

Figure  \ref{CNdistriMass} compares the effect of thermohaline mixing on the surface abundances of red clump stars with (blue) and without (grey) extra mixing, for 1 M$_{\odot}$ (upper panel) and 2 M$_{\odot}$ (lower panel). On the main sequence, low-mass stars (LMS, M$\lesssim$1.7M$_\odot$) burn hydrogen mainly through  pp-chains rather than CNO cycle as in intermediate-mass stars (IMS, 1.7$\lesssim$M$\lesssim$2.2M$_\odot$). Consequently, a large production of $^3$He occurs in LMS, favouring the development of thermohaline instability during the red giant branch. Then the impact on [C/N] distribution is larger for 1.0~M$_{\odot}$ stars than for 2.0~M$_{\odot}$ stars.\\
On the other hand, the RGB-bump occurs on the evolution of low- and intermediate-mass stars only, and depends on the metallicity of these stars. More massive stars (HMS, M$\gtrsim$2.2M$_\odot$) ignite central helium burning in a non-degenerate core at relatively low luminosity on the RGB, well before the hydrogen burning shell reaches the mean molecular weight discontinuity caused by the first dredge-up. Consequently, these objects do not go through the bump on their short ascent of the RGB, and thus thermohaline instability does not develop in this kind of star. 

\subsection{Impact as a function of metallicity}

In low-mass, low-metallicity giants the thermohaline unstable region is more compact and has a steeper temperature gradient, resulting in a higher diffusion coefficient and then a more efficient transport process \citep[see Fig. 6 of ][]{Lagarde11}. 

   \begin{figure}
   \centering
      \textbf{[Fe/H] $\sim 0$}\par
      \includegraphics[width=8cm,clip=true,trim= 0cm 0cm 0cm 0cm]
   {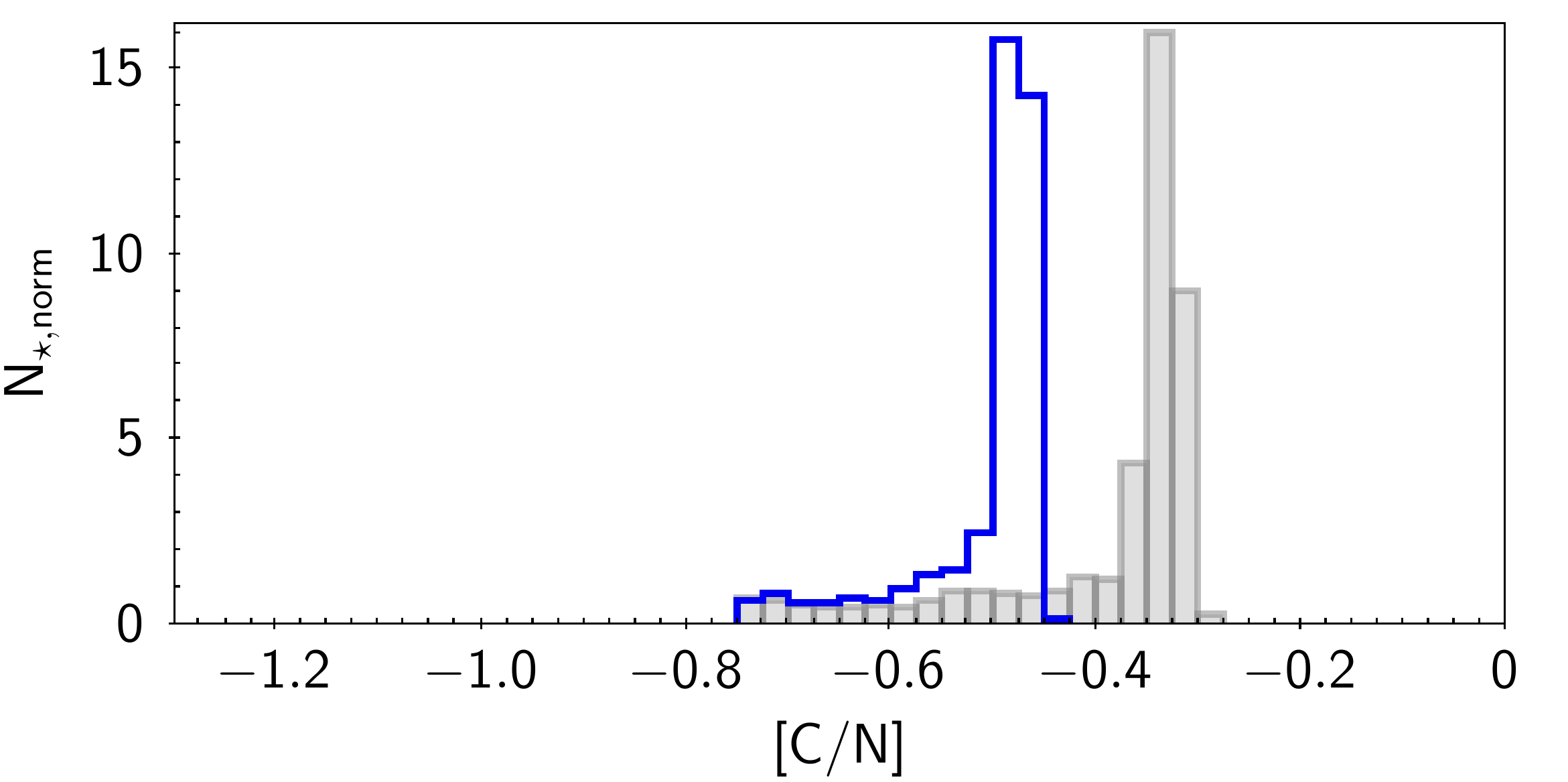}
      \par\vspace{0.3cm}
       \textbf{[Fe/H] $\sim -0.8$}\par
   \includegraphics[width=8cm,clip=true,trim= 0cm 0cm 0cm 0cm]
   {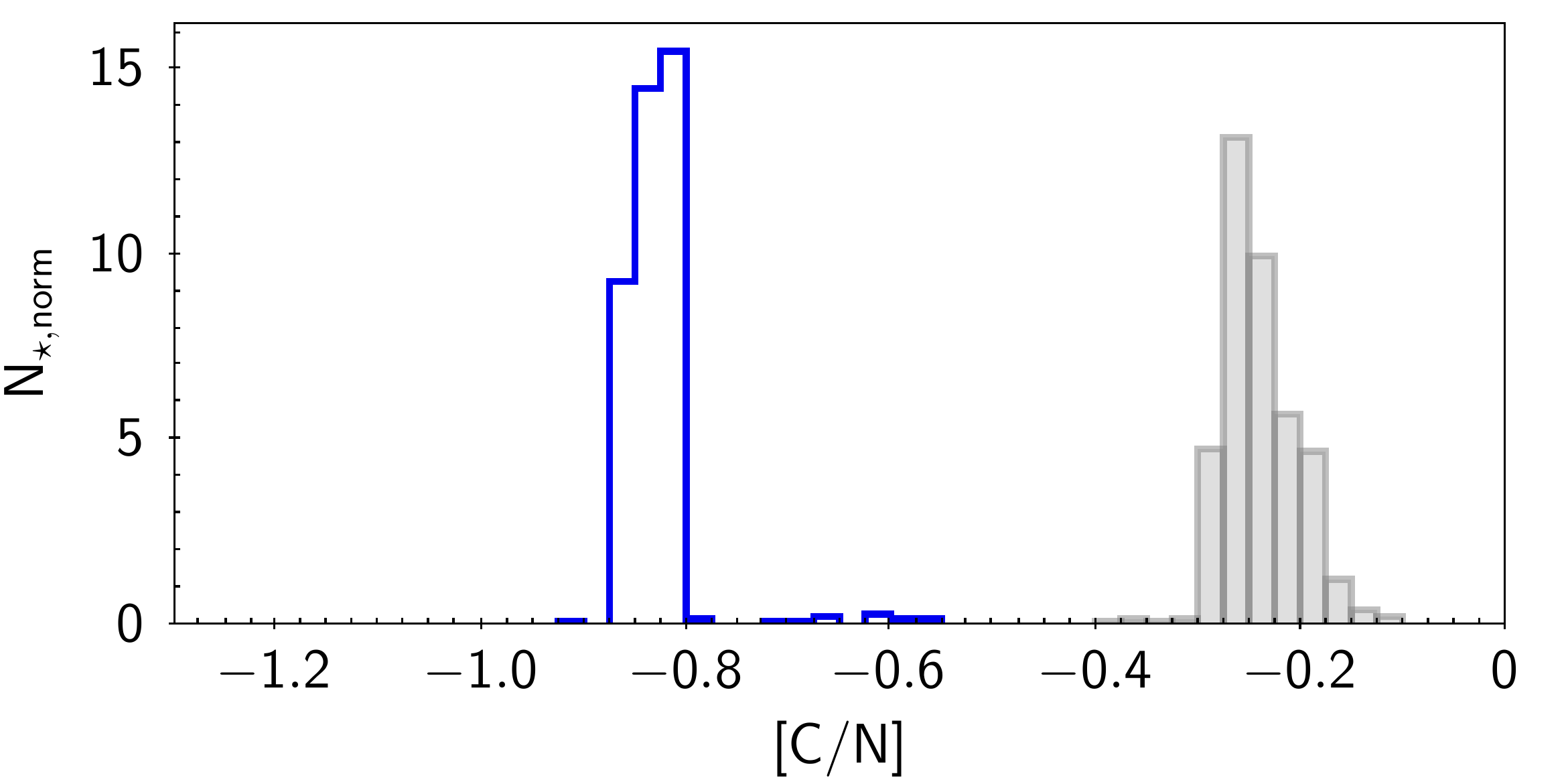}
      \par\vspace{0.5cm}
    \includegraphics[width=8cm,clip=true,trim= 0cm 0cm 0cm 0cm]
   {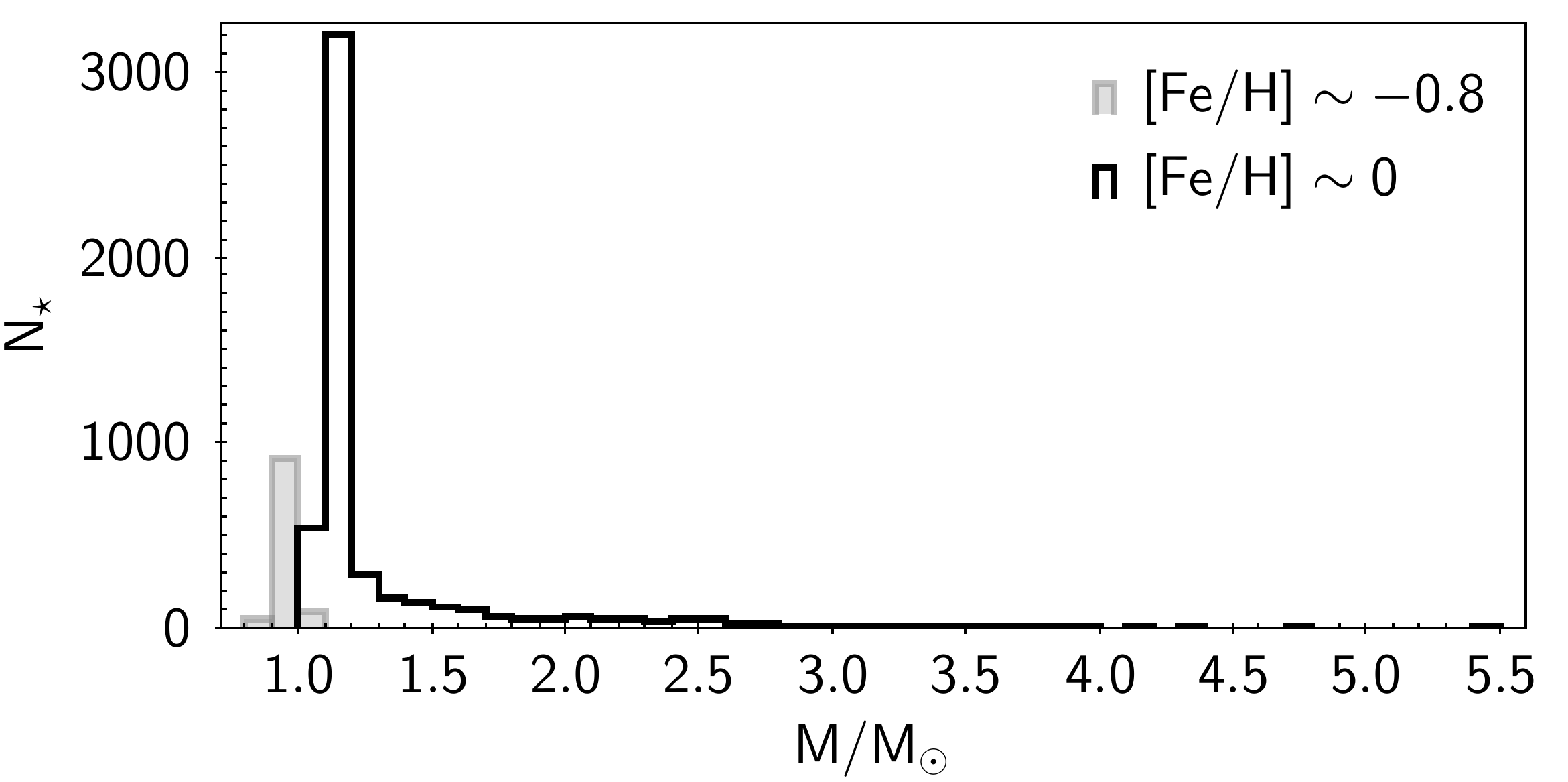} 
      \caption{[C/N] distributions at two metallicities (top and middle panel, respectively) for clump stars simulated with the BGM with the effects of thermohaline instability (blue solid histograms) and without (grey shaded histograms). The bottom panel presents the mass distributions for clump stars simulated with the BGM at -0.85$\leq$[Fe/H]$\leq$-0.75 and -0.05$\leq$[Fe/H]$\leq$0.05.}
         \label{CNdistriZ}
   \end{figure} 

The effect of thermohaline mixing at two metallicities on the [C/N] value at the surface of clump stars simulated by the BGM is shown on Fig. \ref{CNdistriZ} (top and middle panels). The mass distributions for each metallicity range are also shown in Fig. \ref{CNdistriZ} (bottom panel).  Although thermohaline mixing has a larger impact on [C/N] when the metallicity decreases, the figure also clearly shows a non-negligible impact at solar metallicity. \\

      \begin{figure*}
   \centering
  \includegraphics[width=0.48\hsize]
   {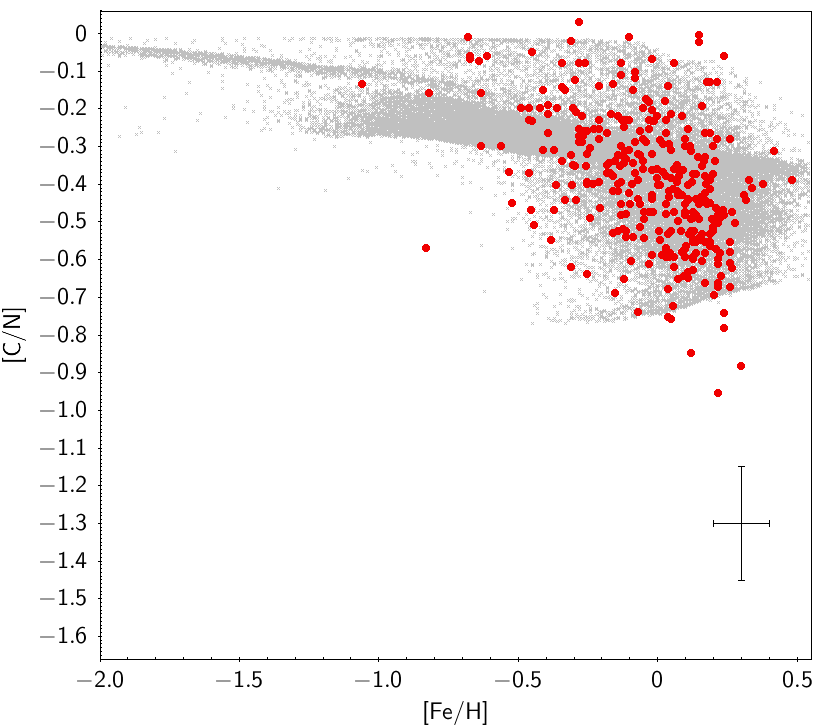}
   \includegraphics[width=0.48\hsize]
   {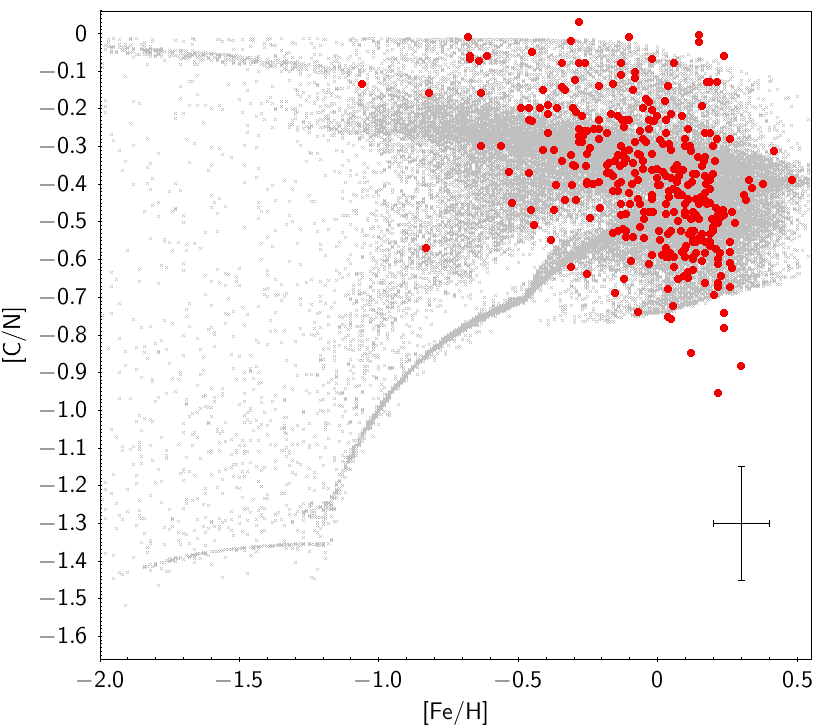}
   \caption{[C/N] as a function of [Fe/H] for synthetic populations computed with the BGM with the effects of thermohaline instability (right panel) and without (left panel). The [C/N] values for our sample of UVES giant field stars are also shown (red dots). }
   \label{CNfield}
 \end{figure*}

Because of their large mass and metallicity ranges, field stars provide key data to constrain the efficiency of thermohaline instability occurring in giant stars. In the next section, we  study the effects of thermohaline instability with mass and metallicity, comparing the population synthesis and the GES observations. \\


\section{Comparison of observed and simulated chemical properties}
\label{field}

 \subsection{C and N abundances in field stars}

 Figure \ref{CNfield} displays the surface [C/N]  of giant stars simulated with the BGM (grey dots) as a function of stellar metallicity, with and without the effects of thermohaline mixing (right and left panel, respectively). Low-RGB,
 upper-RGB, and clump stars are included in this figure (as described above). This figure clearly shows a stronger impact of thermohaline instability on the surface abundances of giants (here, [C/N])  with [Fe/H]<-0.5. Lower metallicity populations are composed essentially of low-mass stars, implying thick disc and halo as key populations to test the efficiency of this extra mixing (see Figs. \ref{CNdistriZ} and \ref{CNfield}).\\

\begin{figure*}
      \hspace{5.5cm}\textbf{Field stars} \hspace{7.5cm} \textbf{NGC 104} \par
      \centering
  \includegraphics[width=0.33\hsize]
   {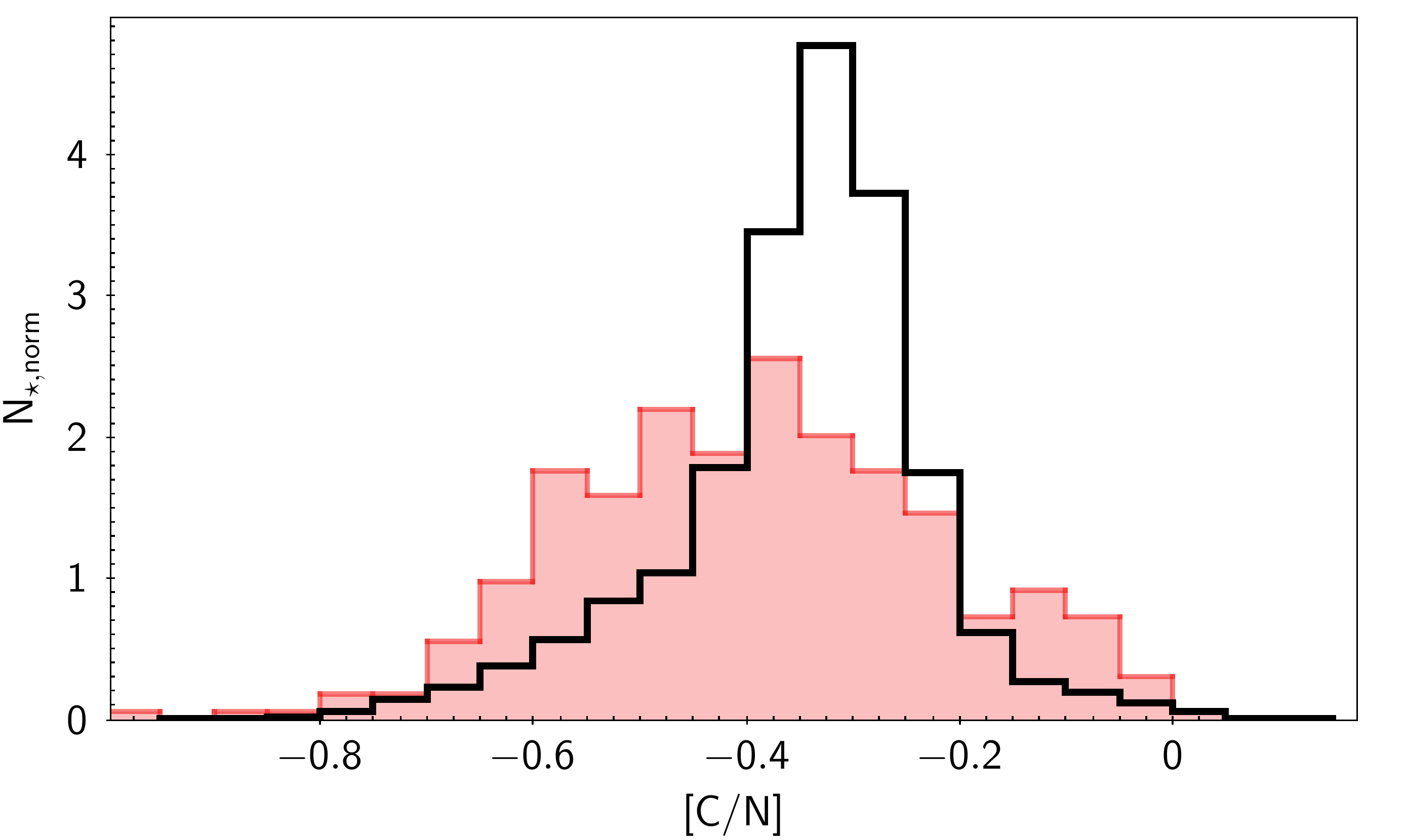}
   \includegraphics[width=0.33\hsize]
   {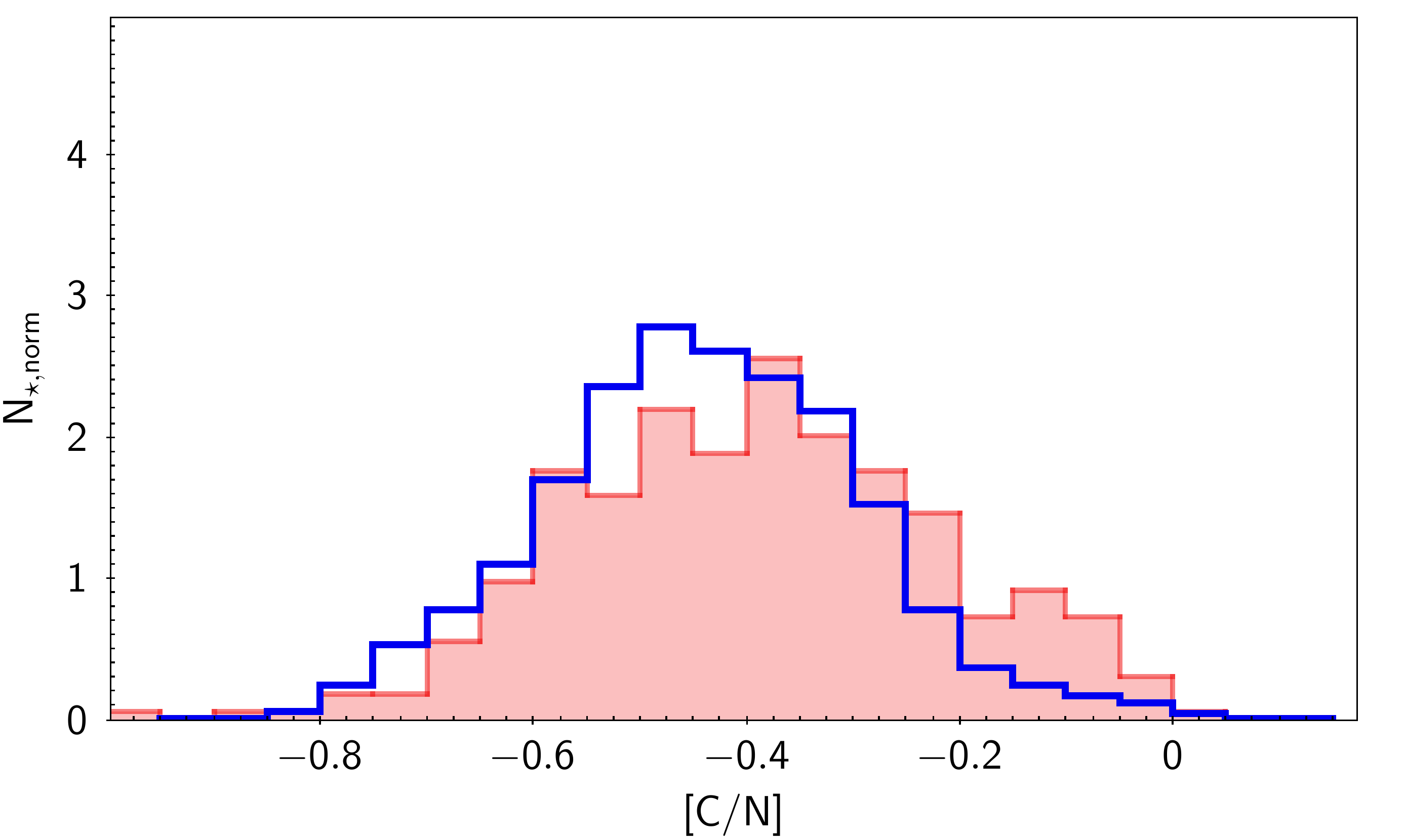}
   \includegraphics[width=0.33\hsize]
   {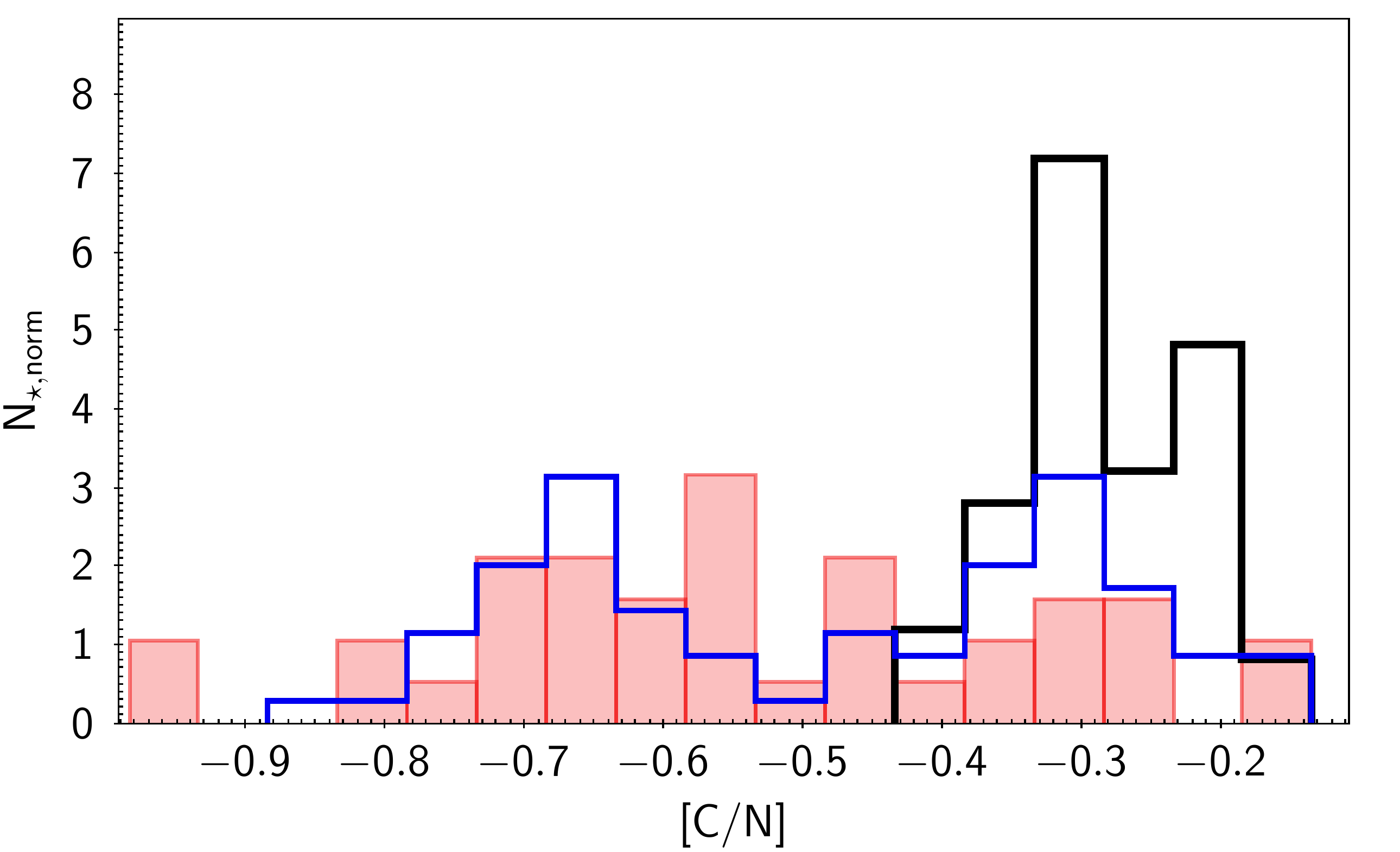}
   \caption{\textit{Left and middle panels:} [C/N] distributions for the  synthetic populations in Fig. \ref{CNfield} with the effects of thermohaline instability (middle panel) and without (left panel). [C/N]  for our sample of UVES giant field stars are also shown (red histogram). \textit{Right panel}: [C/N] distributions for a synthetic populations computed with the BGM for the globular cluster NGC104 with the effect of thermohaline instability (blue histogram) and without (black histogram). The observed [C/N]  derived by GES survey is shown (red histogram).}
   \label{CNfield_distribution}
 \end{figure*}

In the same figure, the [C/N] value derived from the observed field stars are also shown (red dots). Since UVES is centred on solar neighbourhood MSTO stars (plus bulge, see Fig.\ref{planis}), where approximately solar metallicity is expected, the metallicity range of observations is around the solar metallicity (\citet[][]{Stonkute16}, -0.5$\lesssim$[Fe/H]$\lesssim$0.5).  Figure\ref{CNfield_distribution} presents a comparison between the [C/N] distribution predicted by the BGM with and without the effects of themohaline instability and the observed distribution of [C/N] in the GES field stars sample. 
Since the [C/N] range in the simulation including the effects of thermohaline instability or following the standard theory are the same, Fig. \ref{CNfield} does not allow us to discriminate between both prescriptions directly. However, Fig. \ref{CNfield_distribution} allows a quantitative comparison and shows the necessity of extra mixing to reproduce the observations even at higher metallicities. 
In this case, information on stellar masses and evolutionary states of field stars are required to add more constraints on the efficiency of extra mixing with stellar mass. This is now possible with asteroseismology, which can be combined with astrometry from Gaia \citep{GaiaDR1}, and will be discussed in a forthcoming paper. 
 
In addition, the right panel of Fig. \ref{CNfield_distribution} shows the simulation performed specifically for NGC104 (with and without extra mixing), focussing on cluster region  in the sky and the specific metallicity range. The mean [C/N]  and the standard deviation predicted by the simulations including extra mixing ($<$[C/N]$>=-0.48 \pm 0.19$) are in better agreement with observations in NGC104 ($<$[C/N]$>=-0.53 \pm 0.20$), than for simulations following the standard stellar evolution model ($<$[C/N]$>=-0.28 \pm 0.05$). With this very promising result, we  focus on clusters observed by the GES survey in the next section.

\subsection{C and N abundances in clusters}
\label{OCGC}
  \begin{figure*}
   \centering
  \includegraphics[width=0.48\hsize]
   {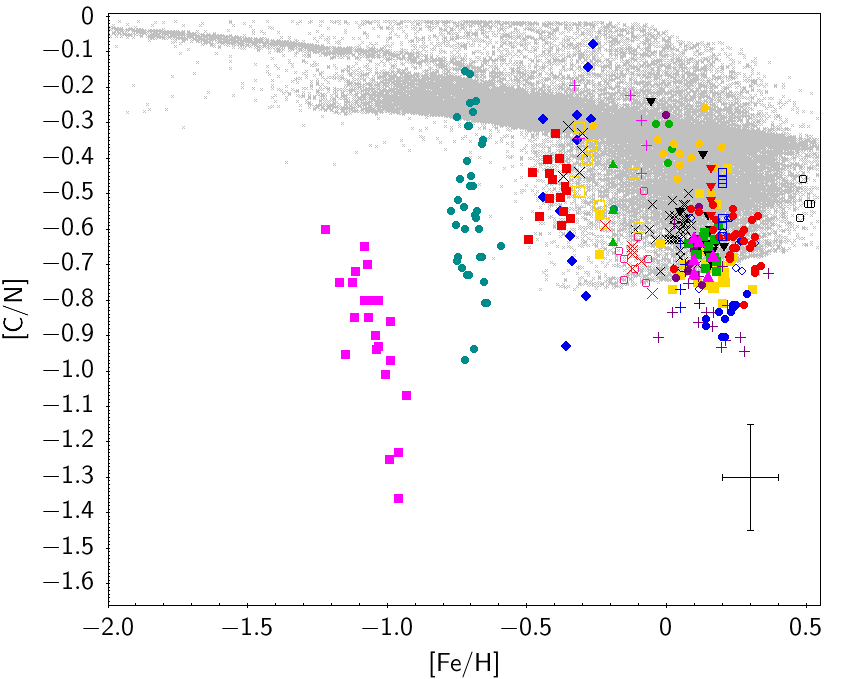}
   \includegraphics[width=0.48\hsize]
   {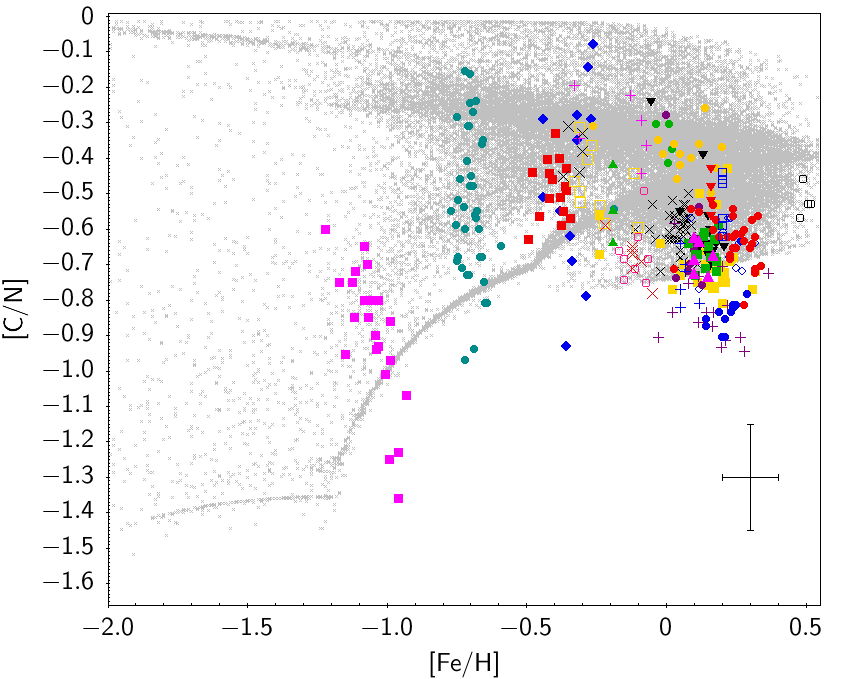}
   \caption{[C/N] as a function of [Fe/H] for synthetic populations computed with the BGM with the effects of thermohaline instability (right panel) and without (left panel). [C/N]  for our sample of UVES giant stars members of open and globular clusters are also shown (each symbol represents a cluster, see Table \ref{clusters}). A typical error bar is indicated. }
   \label{clusters_CN_Feh}
 \end{figure*}

   \begin{figure*}
   \centering
  \includegraphics[width=0.48\hsize]
   {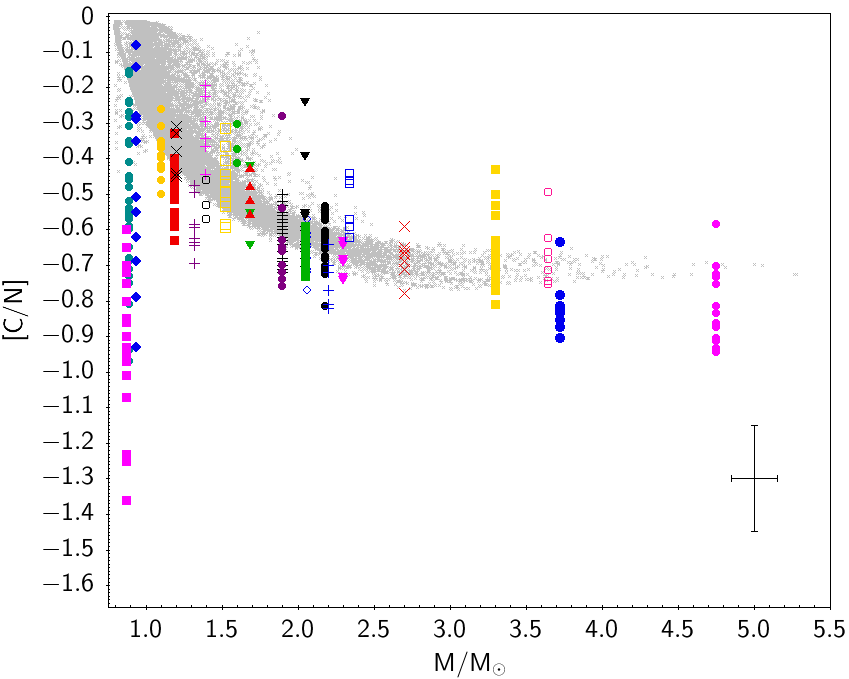}
   \includegraphics[width=0.48\hsize]
   {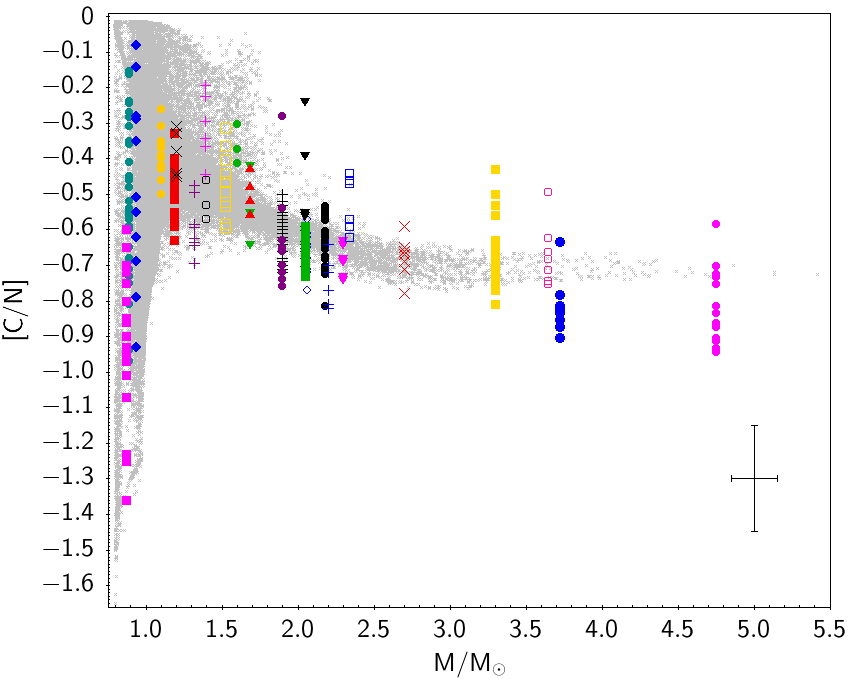}
  \caption{[C/N] as a function of stellar mass for synthetic populations computed with the BGM with the effects of thermohaline instability (right panel) and without (left panel). [C/N]  for our sample of UVES giant stars members of open and globular clusters are also shown (each symbol represents a cluster, see Table \ref{clusters}). A typical error bar is indicated.}
  \label{clusters_CN_Mass}
 \end{figure*}

The GES has observed many different clusters in different regions of our Galaxy (see red circles in Fig. \ref{planis}), providing the homogeneous observational data needed to constrain stellar and Galactic evolution. We investigate the [C/N] value derived in giant members of those open and globular clusters.  We do not want to study each cluster in detail; instead,  clusters are used here as tracers of extra mixing. Since stars belonging to a cluster were formed together, we can assume that they have the same age, distance, and metallicity, resulting in stronger constraints of thermohaline efficiency.\\
 
\begin{figure*}
 \centering
  \includegraphics[width=0.48\hsize]
   {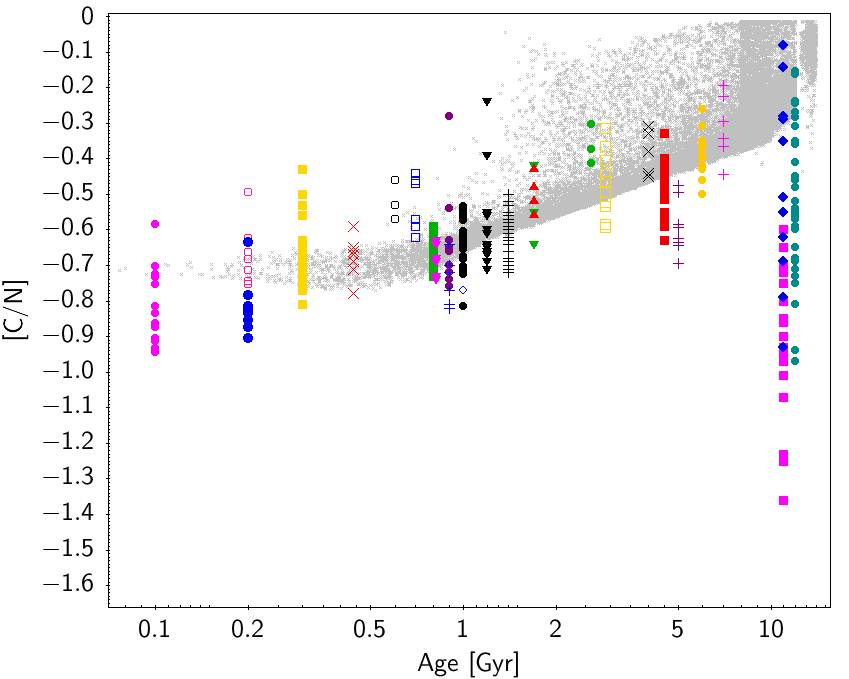}
   \includegraphics[width=0.48\hsize]
   {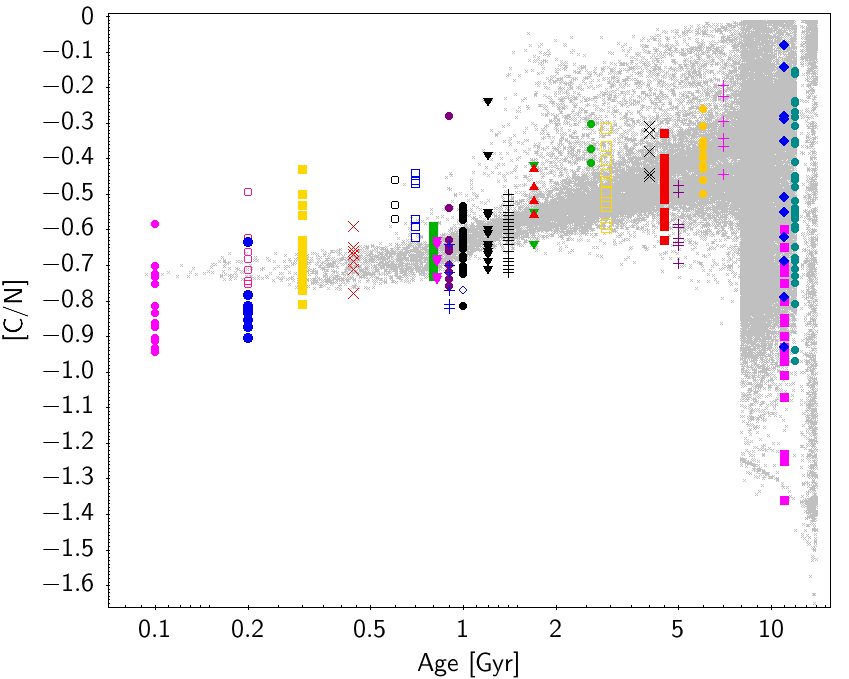}
   \caption{[C/N] as a function of stellar ages for synthetic populations computed with the BGM with the effects of thermohaline instability (right panel) and without (left panel). [C/N]  for our sample of UVES giant stars members of open and globular clusters are also shown (each symbol represents a cluster, see Table \ref{clusters}).}
   \label{clusters_CN_Age}
 \end{figure*}

Figures  \ref{clusters_CN_Feh}, \ref{clusters_CN_Mass}, and  \ref{clusters_CN_Age} show the [C/N] value derived by GES for evolved stars belonging to different globular and open clusters as a function of metallicity, turn-off mass, and age. We also add clusters for which [C/N] and $^{12}$C/$^{13}$C determinations are available from the literature (not from GES): Collinder 261, Melotte 66,  NGC 6253, NGC3960, NGC2324, NGC 2477, NGC 2506, IC 4651, NGC 6134 \citep{Mikolaitis12,Drazdauskas16, Tautvaisiene16,Mikolaitis11a,Mikolaitis11b,Mikolaitis10}. These clusters were acquired by two complementary programmes, and were analysed in a homogeneous way by the same group that produces C and N determinations in GES, resulting in  a robust comparison with our models. Individual stars are attributed the turn-off mass and age of their host clusters (e.g. \citealt{Vandenberg13}, Chantereau's private communication using stellar models discussed in \citealt{ChCh16}, and Drazdauskas et al. in prep.).
Synthetic populations with and without  the effects of thermohaline instability are shown (right and left panels, respectively) for low-RGB, upper-RGB, and clump stars (as defined in Sect. \ref{field}). Simulations and observations in each cluster are a mix of giant stars at different evolutionary states (i.e. at different luminosities or gravities on the RGB), implying a wide [C/N] range for the synthetic populations and for the determination in each cluster.\\
 
As discussed by \citet{ChaLag10} and described in Sect. \ref{cn}, three mass ranges can be evoked to quantify the efficiency of thermohaline instability: 
\begin{itemize}
\item For low-mass and low-metallicity stars, thermohaline instability is the most efficient transport process that can change the C and N surface abundances, which is why the simulations presented here  (which take into account thermohaline mixing only)  reproduce very well the observed [C/N]  in this mass range (see Fig. \ref{clusters_CN_Mass}) including a very good fit for stars  older than $\sim$1 Gyr (see Fig. \ref{clusters_CN_Age}).\\

\item For intermediate-mass stars (1.7$\lesssim$M$\lesssim$2.2M$_\odot$), our simulations present a slightly higher [C/N]  than observations (see right panel of Fig. \ref{clusters_CN_Mass}). \citet{ChaLag10} show that in addition to thermohaline mixing, rotation-induced mixing plays an equivalent role in changing the surface abundances of the stars in this mass range, resulting in a slightly lower [C/N]  at the surface of intermediate-mass stars (see Fig. 17 of \citealt{ChaLag10}). \\

\item For high-mass stars, thermohaline mixing plays no role because these stars do not go through the RGB-bump on their short first ascent of the red giant branch, and thus thermohaline instability does not occur (see Sect. \ref{cn}). This  explains why our simulations do not reproduce the spread of [C/N]  observed in clusters more massive than 2.2M$_\odot$ (see Fig. \ref{clusters_CN_Mass}) and younger than 0.5 Gyr (see Fig. \ref{clusters_CN_Age}). As known for a long time \citep[e.g.][]{MeMa02,Palacios06,ChaLag10}, rotation-induced mixing has an impact on the internal chemical structure of main sequence stars, although its signatures are revealed  later, at the beginning of the RGB. This results in a decrease in the surface abundances of C while N increases. We plan to focus on the effects of rotation on stellar ages and chemical properties in a  forthcoming paper. 
 
 \end{itemize}
 
We note that using the Data Release 12 of the APOGEE survey, \citet{Masseron15} showed that extra mixing has occurred in thin disc stars, but indicated that thick disc stars do not show any evidence of  this extra mixing process. They proposed that the thick disc stars could be formed with a different initial abundances than thin disc stars. We also note that our simulations reproduce very well observations in clusters having metallicity corresponding to the thick disc or halo population assuming the same initial abundances for all populations. However, our comparison is based on clusters members, and complementary to this study it is crucial to consider a larger sample of thick disc field stars such as those observed by the APOGEE survey. In addition, the chemical evolution model and the population synthesis model should be combined to study the effects of different initial abundances for the different populations before drawing any  conclusions. 

\subsection{$^{12}$C/$^{13}$C in field stars and clusters}
\label{c1213_para}

 \begin{figure}
   \centering
   \includegraphics[width=0.99\hsize]
 {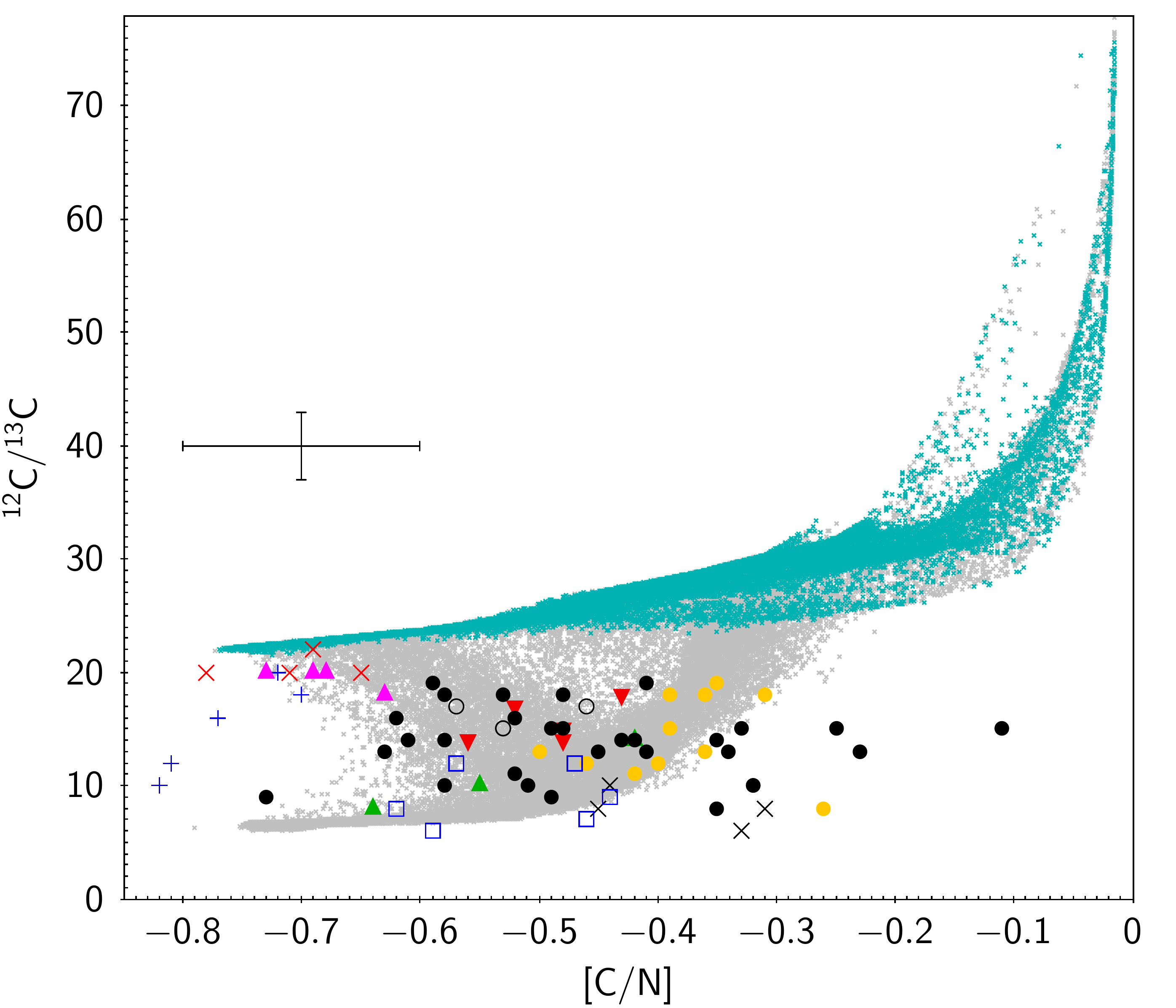}
   \caption{$^{12}$C/$^{13}$C as a function of [C/N]  for synthetic populations computed with the BGM with the effects of thermohaline instability (grey dots) and without (green dots) for upper-RGB and clump stars. A sample of clump field stars are represented by black dots in the right panel, and nine clusters (Collinder 261, Melotte 66, NGC6253, NGC3960, NGC2324, NGC2477, NGC2506, IC4651, NGC6134) using different colours and symbols (see Table \ref{clusters}).}
   \label{c1213}
 \end{figure}

 \begin{figure*}
   \centering
     \includegraphics[width=0.48\hsize]{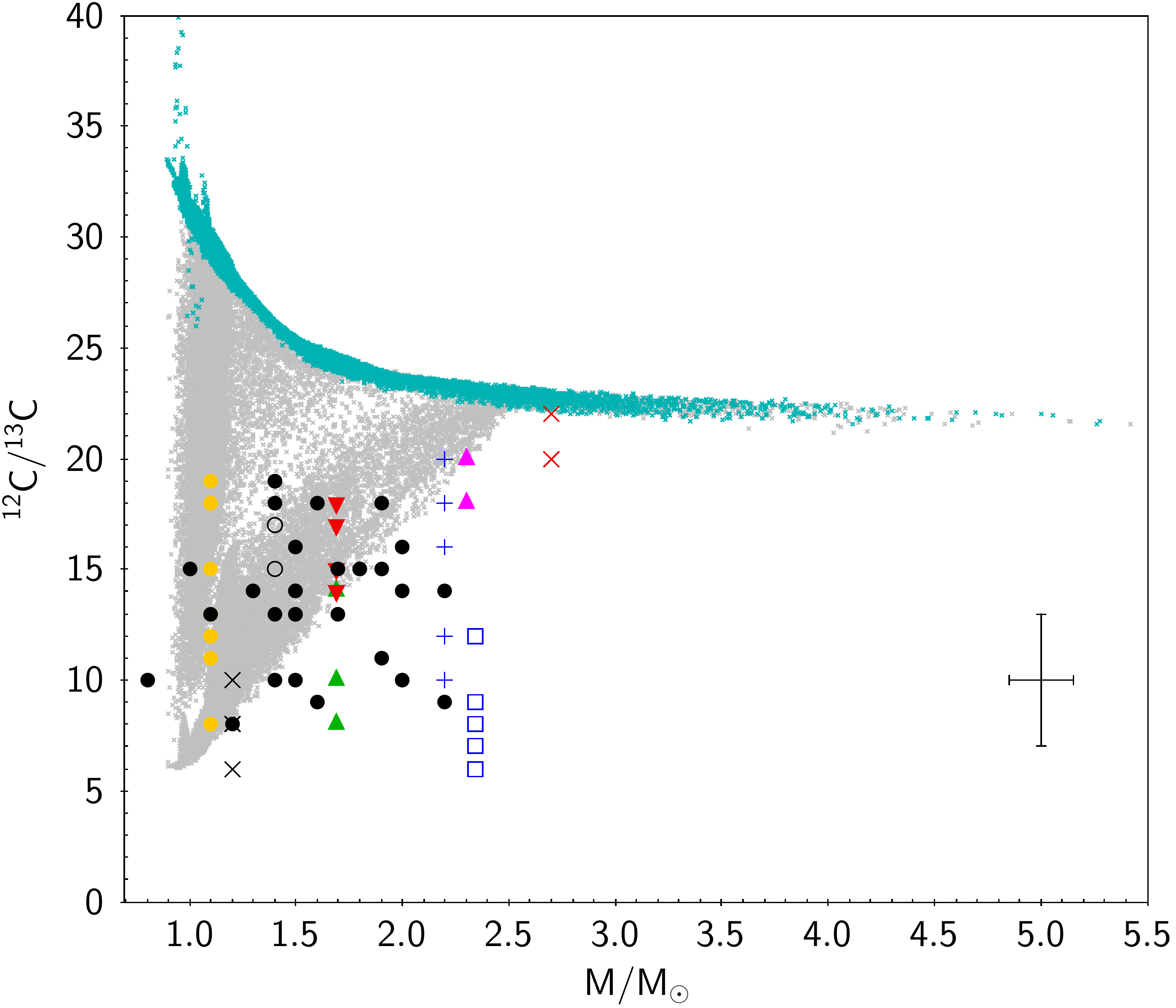}
     \includegraphics[width=0.48\hsize]{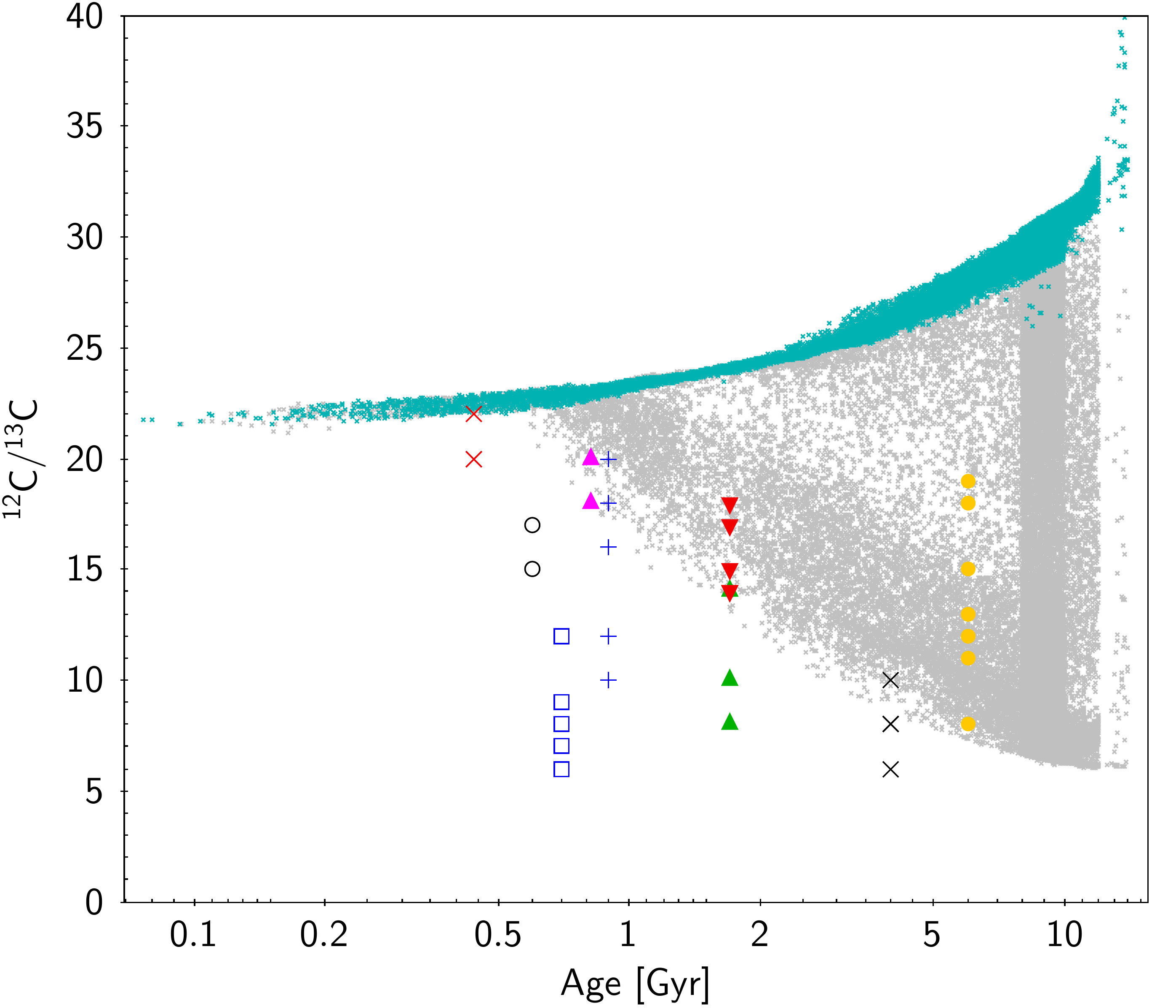}
   \caption{$^{12}$C/$^{13}$C as a function of stellar mass and  ages for synthetic populations computed with the BGM with the effects of thermohaline instability (grey dots) and without (green dots). A sample of clump field stars are represented by black dots in the right panel, and nine clusters (Collinder 261, Melotte 66, NGC6253, NGC3960, NGC2324, NGC2477, NGC2506, IC4651, NGC6134) using different colours and symbols (see Table \ref{clusters}).}
   \label{c1213_all}
 \end{figure*}

The GES cannot derive the carbon isotopic ratio due to the
wavelength regions observed with no good $^{13}$CN features, thus in this part we use data from other studies to investigate the importance of $^{12}$C/$^{13}$C to constrain extra mixing on the red giant branch.
 
Figures  \ref{c1213} and \ref{c1213_all} present the carbon isotopic ratio as a function of C/N and as function of stellar masses and ages, for synthetic populations computed with the BGM taking into account the effects of thermohaline mixing (grey dots) or not (green dots). 
The carbon isotopic ratio decreases abruptly when the thermohaline mixing develops in RGB stars, as already shown by \citet{ChaLag10}. This is in 
agreement with the observed abundance ratios found in field stars and open clusters, shown with symbols in the figures \citep{Mikolaitis12,Drazdauskas16, Tautvaisiene16}.  Even though the range of [C/N] values agrees with both models in the observed range, the low values of $^{12}$C/$^{13}$C cannot be reproduced without extra mixing process.  As discussed below in the [C/N] case, thermohaline mixing explains very well the low-mass (and older) giants stars, but not the higher mass stars. This proves 
that $^{12}$C/$^{13}$C is a more powerful parameter for  constraining extra mixing on the RGB than [C/N], including for solar metallicity stars. 
\begin{table*}
\caption{References for the chemical properties of globular and open clusters used in the comparison with model predictions. For GES clusters, the approximate turn-off masses were evaluated using the theoretical PARSEC isochrones \citep{Bressan12b} with the corresponding ages and metallicities. For globular clusters see  the text;  for clusters already published, we use the turn-off mass and age indicated in the corresponding article.
}        
\begin{center}
\scalebox{0.9}{
\label{clusters}      
\centering                          
\begin{tabular}{|c |c| c| c| c | c|}        
\hline\hline                 
\textbf{Cluster} & \textbf{M$_{TO}$} & \textbf{Age} & \textbf{Ref. Mass \& Age}
& \textbf{Ref. [C/N] } &\textbf{Symbols } \\ 
\hline               
\hline         
Tr 20 & 1.9  & 1.4  & \citet{Carraro10,Donati14}
& GES &  black plus sign \\
\hline
NGC 104 & 0.89 & 12  & \citet{ChCh16}
& GES & dark cyan circle  \\
&  &  & according to \citet{Vandenberg13} and \citet{Parada16}
&    &     \\
\hline
NGC 1851 & 0.87 & 11 & \citet{ChCh16}
& GES & magenta square  \\ 
&  &  & according to \citet{Vandenberg13}
&    &     \\
\hline
NGC 5927 & 0.94 & 11 & \citet{ChCh16}
& GES  & blue diamond  \\
&  &  & according to \citet{Vandenberg13}
&    &     \\
\hline
NGC 6705 & 3.30 & 0.3 &\citet{Santos05}
& GES & yellow square  \\
 & & & \citet{Cantat14}& & \\
\hline
NGC 2243 & 1.19 & 4.5  & WEBDA database
& GES & red square  \\
\hline
Br 81 & 2.06 & 1.0  &\citet{Sagar98}
& GES  &  blue open diamond \\
\hline
NGC 6005 &2.05  &1.2   &\citet{Piatti98}
&GES    & black down triangle \\
\hline
NGC 6802&1.90  & 0.9 & \citet{Tang17}
&GES    & purple circle \\
\hline
Tr 23  & 2.05 & 0.8 &\citet{Overbeek17}   
& GES    & green square  \\
\hline
Br 31& 1.52 & 2.9 & \citet{Cigoni11}
&  GES   & yellow open square \\
\hline
Br 36 &  1.39 & 7 & \citet{Donati12}
& GES   & magenta plus sign \\
\hline
Melotte 71& 3.65 &0.2   & WEBDA database
& GES   & pink open circle  \\
\hline
NGC 6067 & 4.75 & 0.1 & WEBDA database 
&GES    & magenta circle  \\
 & & &according to  \citet[][]{Alonso17} & & \\
\hline
NGC6253  & 1.32 & 3-5 & WEBDA database
& GES    & purple plus sign \\
\hline
M67  & 1.6 &  2.6&WEBDA database
&  GES   & green circle \\
\hline
NGC6259 & 3.73 & 0.2 & WEBDA database
&  GES   & dark blue circle  \\
 &  &  & \citet{Dias02}
&    &     \\
\hline
Rup 134 & 2.18 & 1.0 &\citet{Carraro06}
&  GES   & red circle \\
\hline\hline
 \multicolumn{6}{c|}{}\\
\hline\hline
NGC 2324 & 2.7 & 0.44  
&\multicolumn{2}{c|}{\citet{Tautvaisiene16}}  & red cross  \\
\hline
NGC 3960  & 2.2  &  0.9 
& \multicolumn{2}{c|}{\citet{Tautvaisiene16}}& blue plus sign  \\
\hline
NGC 6253  & 1.4 & 0.6 
&\multicolumn{2}{c|}{\citet{Mikolaitis12}}  & black open circle \\
\hline
NGC  2477 & 2.3 & 0.82  
& \multicolumn{2}{c|}{\citet{Tautvaisiene16}} &  magenta up triangle \\
\hline
Melotte 66  & 1.2 & 4  
&\multicolumn{2}{c|}{\citet{Drazdauskas16} } &  black cross \\
\hline
Collinder 261 & 1.1 & 6.0   
&\multicolumn{2}{c|}{\citet{Mikolaitis12}} & yellow circle \\
  & & 
  &\multicolumn{2}{c|}{\citet{Drazdauskas16}}& \\ 
\hline
NGC 6134 & 2.34 & 0.7 
& \multicolumn{2}{c|}{\citet{Mikolaitis10}} & blue open square\\
\hline
NGC 2506 & 1.69 & 1.7 
& \multicolumn{2}{c|}{\citet{Mikolaitis11a}} & green up triangle\\
\hline
IC 4651 & 1.69 & 1.7 
& \multicolumn{2}{c|}{\citet{Mikolaitis11b}} & red down triangle  \\
  \hline
\end{tabular}}
\label{clustertable}
  \end{center}
\end{table*}

\section{Conclusions}
\label{conclu}

In this paper, we present the first comparison between synthetic populations computed with the Besan\c con Galaxy model and the C and N abundances derived by the Gaia-ESO survey in field stars and in different clusters, both globular and open. We conclude from this work  that it is crucial to take into account thermohaline mixing to understand the C and N observed at the surface of low-mass stars, and in the determination of the stellar mass and age from their chemical properties \citep{Martig15,Ness16}.  \\

We compared data from the Gaia-ESO survey with predictions computed using the Besan\c con Galaxy model in which we included stellar evolution models taking  into account (and not) the effects of thermohaline instability (Paper I). To date, this mixing is the only physical process proposed in the literature to explain the photospheric composition of evolved red giant stars. We focus in the first part of this paper on field stars because of their wide-coverage properties (e.g. mass, metallicity, and ages) to deduce an observational trend between [C/N]  and stellar mass, metallicity, or age.  Due to the lack of C and N determinations in field stars at low metallicity, we cannot investigate further the observational constraints in the metallicity domain where thermohaline instability is more efficient. Nevertheless, the theoretical distribution of [C/N] predicted by the BGM including the effects of thermohaline instability is in better agreement than the distribution predicted by the standard model at metallicity close to solar.\\

We also investigate the [C/N]  derived in giant members of open and globular clusters by the Gaia-ESO survey and literature. This comparison shows a very good agreement with stellar evolution models including thermohaline mixing over the whole scrutinized metallicity range,  explaining the [C/N]  observed in lower mass and older giant stars. This confirms that thermohaline instability is crucial to understand the chemical properties of giant stars. The next step is to use a larger sample of field stars to strengthen this encouraging result. \\
On the other hand, we show that the observed behaviour of $^{12}$C/$^{13}$C with  stellar ages is clearly reproduced by models which include the effect of thermohaline mixing. This confirms the importance, amongst others, of extra mixing when deducing  stellar ages from the chemical properties of giant stars.\\

Additionally, and independently of spectroscopy, asteroseismology paves the way to a better understanding of stellar interiors, providing valuable and independent constraints on current stellar evolution models and on the physics of different transport processes. The space missions CoRoT \citep{Baglin06}, \textit{Kepler}, and K2 \citep{Borucki10}  observed a large number of giant stars in different regions in our Galaxy, allowing a unique opportunity to derive some fundamental properties (e.g. stellar mass, radius, age and gravity, and evolutionary stage of giants) by observation of mixed modes in red giants \citep[e.g.][]{ChMi13}. To obtain the most information possible from the data sample, the asteroseismic properties must be combined with the observations of the surface chemical abundances and especially the surface $^{12}$C/$^{13}$C. Future studies of CoRoT inner-field stars (Valentini et al. in prep.) and K2 giants (campaign 3)  already observed by GES, will provide complementary results for the development of stellar evolution models.  

\begin{acknowledgements}
 Based on data products from observations made with ESO Telescopes at the La Silla Paranal Observatory under programme ID 188.B-3002. These data products have been processed by the Cambridge Astronomy Survey Unit (CASU) at the Institute of Astronomy, University of Cambridge, and by the FLAMES/UVES reduction team at INAF/Osservatorio Astrofisico di Arcetri. These data have been obtained from the Gaia-ESO Survey Data Archive, prepared and hosted by the Wide Field Astronomy Unit, Institute for Astronomy, University of Edinburgh, which is funded by the UK Science and Technology Facilities Council.\\
 
This work was partly supported by the European Union FP7 programme through ERC grant number 320360 and by the Leverhulme Trust through grant RPG-2012-541. We acknowledge the support from INAF and Ministero dell'Istruzione, dell'Universit\`a e della Ricerca (MIUR) in the form of the grant `Premiale VLT 2012'. The results presented here benefit from discussions held during the Gaia-ESO workshops and conferences supported by the ESF (European Science Foundation) through the GREAT Research Network Programme.\\
   
N.L., C.R., A.R., and G.N. acknowledge financial support from the `Programme National de Physique Stellaire' (PNPS) and the `Programme National de cosmologie et Galaxie' (PNCG) of CNRS/INSU, France. N.L. acknowledges financial support from the CNES. Simulations have been executed on computers from the Utinam Institute of the Universit\'e de Franche-Comt\'e, supported by the R\'egion de Franche-Comt\'e and Institut des Sciences de l'Univers (INSU). G.T., A.D., \v{S}.M., R.M., E.S., Y.Ch., and V.B acknowledge support from the Research Council of Lithuania (MIP-082/2015). R.S. acknowledges support from the Polish Ministry of Science and Higher Education. F.J.E. acknowledges financial support from ASTERICS project (ID:653477, H2020-EU.1.4.1.1. - Developing new world-class research infrastructures). T.B. was funded by the project grant `The New Milky Way?' from the Knut and Alice Wallenberg Foundation. A.J.K acknowledges support from the Swedish National Space Board (SNSB). T.M. acknowledges support provided by the Spanish Ministry of Economy and Competitiveness (MINECO) under grant AYA-2017-88254-P.
 \end{acknowledgements}

\bibliographystyle{aa}
\bibliography{Reference}
\end{document}